\documentclass[journal=jctcce,manuscript=article]{achemso}
\usepackage[utf8]{inputenc}
\usepackage{graphicx}
\usepackage{amsmath}
\usepackage{caption}
\usepackage[subrefformat=parens]{subcaption}
\graphicspath{ {./images/} }

\usepackage{
 	amssymb,
	float,
	ifthen,
	tikz,
	docmute 
}
\usetikzlibrary{
	cd,
	calc,
	intersections
}

\title{Characterizing Reaction Route Map of Realistic Molecular Reactions based on Weight Rank Clique Filtration of Persistent Homology}
\author{Burai Murayama}
\affiliation[Chem Sci Hokkaido Univ]
{Department of Chemistry, Faculty of Science, Hokkaido University, Sapporo 060-0810, Japan}

\author{Masato Kobayashi}
\affiliation[Chem Sci Hokkaido Univ]
{Department of Chemistry, Faculty of Science, Hokkaido University, Sapporo 060-0810, Japan}
\alsoaffiliation[ICReDD Hokkaido Univ]
{WPI-ICReDD, Hokkaido University, Sapporo 060-0810, Japan}
\email{k-masato@sci.hokudai.ac.jp}

\author{Masamitsu Aoki}
\author{Suguru Ishibashi}
\author{Takuya Saito}
\affiliation[Math Sci Hokkaido Univ]
{Department of Mathematics, Faculty of Science, Hokkaido University, Sapporo 060-0810, Japan}

\author{Takenobu Nakamura}
\affiliation[AIST]
{National Institute of Advanced Industrial Science and Technology, Tsukuba 305-8568, Japan}

\author{Hiroshi Teramoto}
\affiliation[Kansai Univ]
{Faculty of Engineering Science, Kansai University, Suita 564-8680, Japan}

\author{Tetsuya Taketsugu}
\affiliation[Chem Sci Hokkaido Univ]
{Department of Chemistry, Faculty of Science, Hokkaido University, Sapporo 060-0810, Japan}
\alsoaffiliation[ICReDD Hokkaido Univ]
{WPI-ICReDD, Hokkaido University, Sapporo 060-0810, Japan}

\begin{document}

\maketitle

\begin{abstract}
A reaction route map (RRM) constructed using the GRRM program is a collection of elementary reaction pathways, each of which comprises two equilibrium (EQ) geometries and one transition state (TS) geometry connected by an intrinsic reaction coordinate (IRC).
An RRM can be mathematically represented by a graph with weights assigned to both vertices, corresponding to EQs, and edges, corresponding to TSs, representing the corresponding energies.
In this study, we propose a method to extract topological descriptors of a weighted graph representing an RRM based on persistent homology (PH).
The work of Mirth \textit{et al.} [\textit{J. Chem. Phys.} \textbf{2021}, \textit{154}, 114114], in which PH analysis was applied to the $(3N-6)$-dimensional potential energy surface of an $N$ atomic system, is related to the present method, but our method is practically applicable to realistic molecular reactions.
Numerical assessments revealed that our method can extract the same information as the method proposed by Mirth \textit{et al.} for the 0-th and 1-st PHs, except for the death of the 1-st PH.  
In addition, the information obtained from the 0-th PH corresponds to the analysis using the disconnectivity graph.
The results of this study suggest that the descriptors obtained using the proposed method accurately reflect the characteristics of the chemical reactions and/or physicochemical properties of the system.  
\end{abstract}

\section{Introduction}
Reaction pathways are fundamental for expressing chemical reactions in theoretical studies.
Reaction pathways are defined by the potential energy surface (PES), which is a scalar function on $(3N-6)$-variables in the case of an $N$ atomic system. \cite{Wales2003_EnergyLandscape,Mezey1987_PES-book}
The first-order saddle point and local minimum point of the PES represent the transition state (TS) geometry and equilibrium (EQ) geometry, respectively. 
The steepest descent curve connecting two EQ geometries via a TS is known as the intrinsic reaction coordinate (IRC) \cite{Fukui1970_IRC}, which represents an elementary reaction pathway.
As the $(3N-6)$ variables are determined based on the positions of the atoms, the reaction pathways aid in the expression of the configurational changes that occur through the chemical reaction from one EQ to another and facilitate the overall understanding of the mechanism of chemical reactions.

A reaction route map (RRM) is a collection of elementary reaction pathways that characterizes chemical reactions in $N$ atomic chemical systems. \cite{Satoh2020_RMapDB,IQCE_GRRM-GDSP}
Mathematically, an RRM is expressed as a weighted graph with assigned weights corresponding to the energies of the EQs or TSs. 
Weighted graphs representing RRMs can describe the physical properties of chemical reactions, such as the reaction rate or lifetime.
Although visualization of the PES is difficult owing to its high dimensionality, RRMs can be visualized on a plane.
Therefore, RRM represents a reduction in information from the PES, transforming an object that cannot be visualized into one that can \cite{Satoh2015_RMapViewer,Satoh2016_RMapViewer}. 
Therefore, RRMs are widely used as a standard representation method for chemical reactions of $N$ atomic systems. \cite{Satoh2020_RMapDB}

Several algorithms for TS search from an EQ structure have been proposed so far.
The eigenvector following (EF) method, originally proposed by Cerjan and Miller \cite{cerjan1981finding}, follows the gentlest ascent path along the eigenvector of the Hessian with the smallest eigenvalue at the EQ to locate a TS. 
J{\o}rgensen \textit{et al.} \cite{jorgensen1988gradient} used the gradient extremal path, where the gradient is an eigenvector of the Hessian, to connect stationary points.
Quapp \textit{et al.} \cite{quapp1998searching} proposed the reduced gradient following (RGF) method by modifying the gradient extremal following method.
Zimmerman \cite{zimmerman2015single} proposed a single-ended growing string method to locate an approximate TS from a single EQ structure.
Further, several algorithms have been proposed for the automatic and extensive construction of RRMs for complicated molecular systems.\cite{Doye1997_Eigenvector-Following}
The global reaction route mapping (GRRM) strategy of Maeda \textit{et al.} \cite{Maeda2013_GRRM} is a typical algorithm.
The GRRM program comprises two methodologies: anharmonic downward distortion following (ADDF) \cite{Ohno2004_ADDF,Maeda2014_ADDF} and artificial force-induced reaction (AFIR) \cite{Maeda2010_MC-AFIR,Maeda2014_SC-AFIR,Maeda2016_AFIR} methods. 
The ADDF method is based on the downward distortion of the one-dimensional potential energy curve, along with isomerization or dissociative reaction from a potential minimum, with respect to the harmonic potential around the minimum. 
The AFIR method is a more intuitive method that applies an artificial force between two atoms or atomic groups to induces a reaction forcibly. 
By applying the ADDF or AFIR method, the GRRM program automatically identifies the EQ and TS geometries sequentially.
Owing to these features, the GRRM program has been used in nanocluster catalysts \cite{Iwasa2019_Cu13NO} and organic synthesis design (Ref.~\citenum{Hayashi2022_GRRM_pyridine-dearomatization}, with a maximum of 300 EQs and 3964 TSs), and surface chemistry (Ref.~\citenum{Sugiyama2019_CO-on-Pt_GRRM}, with 133 EQs and 298 TSs).
In addition, RRMs can be constructed using other programs such as Chemoton \cite{unsleber2022chemoton, github_chemoton}, KinBot \cite{van2020kinbot,zador2023automated, github_KinBot}, AARON \cite{guan2018aaron, ingman2021qchasm, github_AARON}, ChemTraYzer \cite{dontgen2015automated, ChemTraYzer}, and others \cite{turtscher2022pathfinder,ismail2022graph,unsleber2020exploration}.
RRMs obtained via these programs are often more complicated than those obtained manually owing to the inclusion of multiple chemically equivalent reaction paths with different conformations.
Thus, massive and complicated RRM data are obtained and used for predictive material design.

Utilization of the GRRM strategy introduces a problem in the representation and feature extraction of RRMs.
In general, the layout of the visualization of graphs is arbitrary.\cite{Battista1994_GraphDrawing,Herman2000_GraphVisualization}
In contrast, the GRRM strategy generates a large graph that contains multiple elementary reaction pathways.
In such large graphs, arbitrariness can lead to the visualization of the same graph in different ways, making it difficult to capture the intrinsic features of the graph.
It is difficult to determine whether two different visualizations represent the same graph, particularly for those with a high number of pathways.
Similar problems exist in the visualization and feature extraction of RRMs obtained using the other algorithms.
Although RRM is already a well-reduced representation of PES, further reduction is required for complicated systems to overcome this problem and extract rich information from RRMs.

Several methods have been proposed to address this issue.
Tsutsumi \textit{et al.} \cite{Tsutsumi2018_CMDS-GRRM,Tsutsumi2021_ReSPer,Tsutsumi2022_ReSPer} proposed a reaction-space projector (ReSPer) method to generate reduced dimensional coordinates automatically based on the structural similarity evaluated via the classical multidimensional scaling method.
In addition, the rate-constant matrix contraction (RCMC) method proposed by Sumiya \textit{et al.} \cite{Sumiya2015_GRRM-RCMC}, which was subsequently used to reduce the exploration space in GRRM \cite{Sumiya2019_RCMC-navigation}, can be regarded as a method that summarizes the given RRMs.
In particular, the disconnectivity graph (DG) \cite{Becker1997_Disconnectivity-graph,Wales2005_Disconnectivity-graph} and persistent homology (PH) \cite{petri2013topological,Mirth2021_EL-PH} are generic methods and promising candidates.
DG displays the hierarchy of stable states on a PES using a tree structure.
It is currently the most popular visualization method and has been applied to a large amount of simulation data, irrespective of whether the basis is a realistic molecular system or a toy model. \cite{Wales2003_EnergyLandscape}
However, based on its original definition, DG can only represent the lowest energy transition barrier between two stable states that are not necessarily directly connected.
Hence, it cannot express cyclic pathways, which are frequently found in RRMs, though a related extension has been developed subsequently. \cite{Okushima2007_ConnectivityGraph} 
Recently, PH was used to extract the topological features of a PES by Mirth \textit{et al.}\cite{Mirth2021_EL-PH}
PH is a recent data analysis technique proposed in the field of applied mathematics.
Thus, fewer studies have been conducted on it compared to the case of DG.
As described later, PH can represent cyclic pathways. 
However, the formulation by Mirth \textit{et al.} requires all stationary points of the PES to be located analytically.
Thus, although their work can be regarded as a starting point for understanding the relationship between the reaction pathways and PH, their method cannot be applied directly to RRMs obtained via GRRM.

Herein, we propose a method for applying PH to RRMs, which is applicable to realistic materials obtained via GRRM.
First, we confirmed that the proposed method can extract information consistent with the results of Mirth \textit{et al.} by applying it to the RRM of \textit{n}-pentane, which was investigated in Ref.~\citenum{Mirth2021_EL-PH} based on the model potential. 
Subsequently, the method was applied to RRMs of metal nanoclusters and organic molecular systems, revealing its capability of extracting an RRM's characteristic features, including not only those extracted by DG but also those newly identified by the proposed scheme, such as cyclic reaction pathways.

The rest of this paper is organized as follows.
In Section \ref{sec:Method}, the computational and mathematical methods used are discussed, including a brief introduction to the GRRM program, the application of PH to a weighted graph, and a demonstration of the proposed scheme on a toy model RRM.
The actual applications to the RRMs of metal nanoclusters and organic molecular systems and their interpretations are presented in Section \ref{sec:Results}, followed by concluding remarks in Section \ref{sec:Conclusion}. 

\section{Methods}
\label{sec:Method}
\subsection{RRMs and ADDF/AFIR Methods}

Here, an RRM is defined as a collection of EQ structures of a given atomic constitution along with TS structures (i.e., first-order saddle points) connecting pairs of EQs along IRCs, together with their corresponding energies.
Because the EQs and TSs on each RRM share a certain atomic constitution, the dissociation channels were not considered.
Therefore, an RRM is represented by an undirected graph $G = (V, E)$, with energy weights assigned to both vertices ($w_V(v_i)$, corresponding to the energy of $i$-th EQ) and edges ($w_E(e_i)$, corresponding to the energy of $i$-th TS).
$G$ may contain loops (i.e., elementary identity reactions) and/or multiple edges (i.e., different elementary reactions with identical reactants and products). 
In practice, we constructed the RRM of a certain system using either the ADDF or AFIR method implemented in the GRRM program. \cite{Maeda2018_GRRM17,Maeda2021_GRRM20}.

The ADDF method \cite{Maeda2014_ADDF} follows the anharmonic downward distortion (ADD) of the PES, which is evaluated as the difference between the harmonic potential approximated by the EQ structure and actual potential energy. 
An efficient method for searching for ADD is to introduce an EQ-centered hypersphere drawn with normal coordinates scaled by $1/2$ power of the force constant, over which the energy at the harmonic approximation level is constant.
Because each minimum on a scaled hypersphere surface corresponds to the ADD, the reaction route can be followed along the ADD by increasing the hypersphere radius.
Although the ADDF method is conceptually well-defined, it requires Hessian computation to construct the scaled hypersphere surface at each EQ, limiting its practical application.
Moreover, not all reaction paths can be searched by tracking the ADD paths because of the bifurcation of anharmonic distortion stationary paths \cite{Ebisawa2021_ADDFpath}.

In contrast, the AFIR method \cite{Maeda2016_AFIR} is a more intuitive and practical procedure for constructing RRMs.
In this method, an artificial force is introduced between the system fragments to induce a reaction without an energy barrier.
The reaction path obtained by minimizing the potential energy augmented by the artificial force is called the AFIR path.
Subsequently, the artificial force term is removed, the maximum energy point is adopted as the approximated TS structure, and the actual TS structure is optimized.
In the AFIR method, the applied force is scaled using the parameter $\gamma$, which corresponds to the model collision energy.
In general, TSs with reaction barriers below $\gamma$ can be identified using the AFIR method.

The GRRM program was designed to efficiently search for undiscovered reaction paths (EQs and TSs).
When the program identifies a candidate EQ or TS structure, it determines whether it is chemically equivalent to a previously identified EQ or TS structure. 
Thus, it does not distinguish between nuclear permutation isomers and nuclear inversion isomers, i.e., only one of the two chiral isomers (enantiomers) is included in the list of EQs or TSs.
This restriction can significantly reduce the cost for the exploration of reaction paths. \cite{Ohno-Satohbook}
However, the unmodified RRM network obtained via GRRM does not correspond to the network on $(3N-6)$-dimensional PES---it is contracted using nuclear permutation-inversion symmetry.
Note that there is a method for discriminating between chiral isomers (e.g., see Ref.~\citenum{sobez2020molassembler}).

\subsection{Outline of Persistent Homology for PES}

This section intends to briefly explain the relationship between PH and PES, as in the previous work by Mirth \textit{et al.}
\cite{Mirth2021_EL-PH}
The explanation here is neither devoted to our method nor mathematically rigorous.
However, it provides an intuitive sketch of the PH for the chemical reaction.
The rigorous introduction to the proposed method is presented in the next section.

\begin{figure}[H]
\centering
\includegraphics[keepaspectratio, scale=0.75]{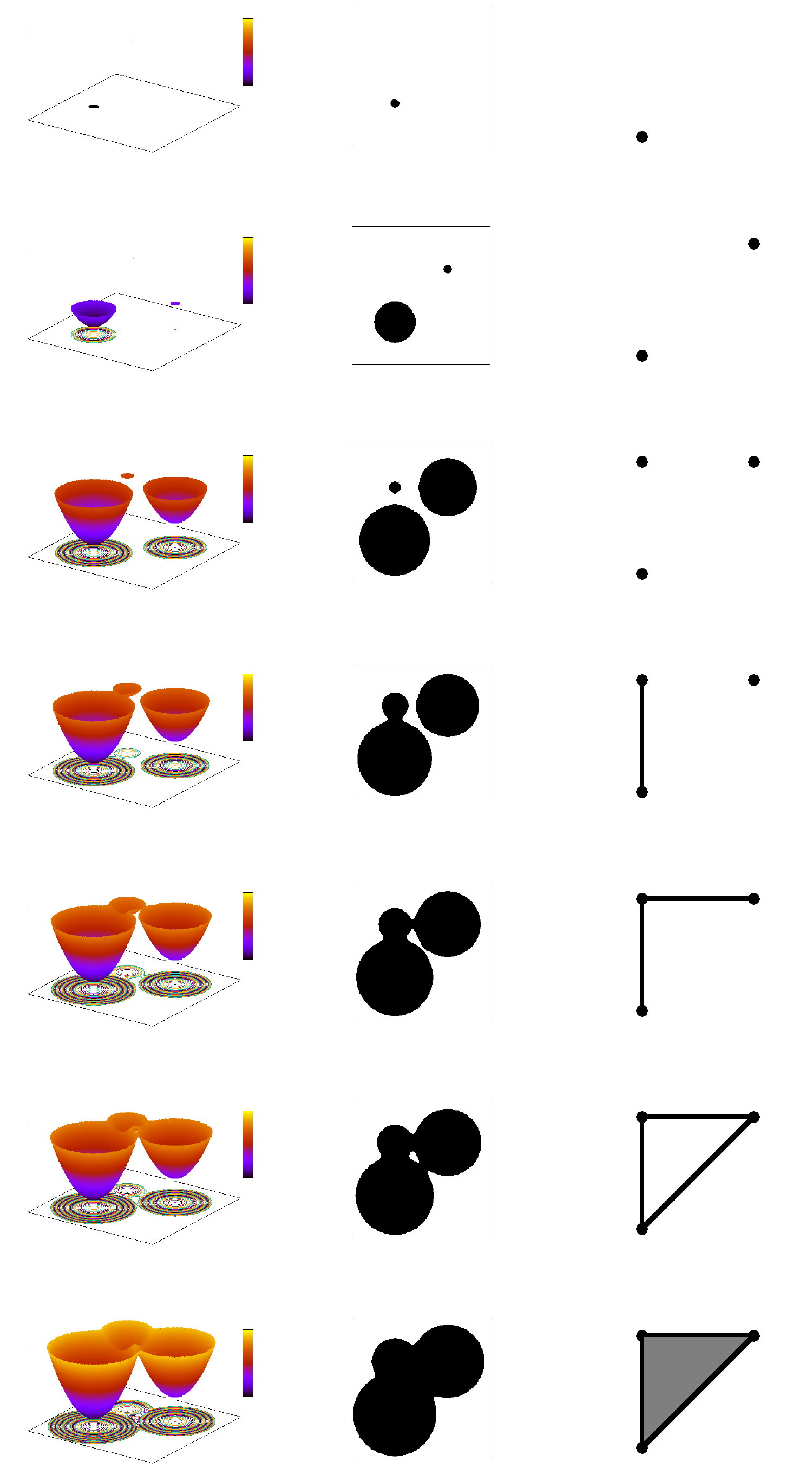}
\caption{Schematic for an explanation of PH for PES.
The lower figure corresponds to greater energy $a$. The PES below $a$ (surfaces in the left column) with the contour map (bases there), $L(a)$ (the center column), and its topological representation (known as simplicial complex) (the right column) are depicted.}
\label{fig:filtration_PES}
\end{figure}

Let $x$ denote a configuration of $3N-6$ component coordinate, and $\Phi(x)$ denote the potential energy.
Suppose the region whose energy is below $a$ i.e. $L(a)=\{x|\Phi(x)\le a\}$.
In other words, for a given topography $\Phi(x)$, $L(a)$ represents the shape of the sea surface with water level $a$.
Figure \ref{fig:filtration_PES} shows a simplified $\Phi(x)$ example introduced for the explanation.
As can be seen in the figure, the shape of the sea surface $L(a)$ itself expands as $a$ increases but retains its topology, except when $a$ crosses a stationary point.

First, when $a$ is below the global minimum of $\Phi(x)$, nothing exists. By increasing $a$ and crossing the minimum, the first basin appears (first row in Fig.~\ref{fig:filtration_PES}). 
Let the value of energy $a$ at this time be the birth $b_0$ of this basin with the label ``$0$''.
As $a$ increases, two basins with labels ``$1$'', ``$2$'' appear when $a$ crosses the local minima (second and third rows in Fig.~\ref{fig:filtration_PES}). 
These minima are recorded as $b_1$ and $b_2$.
Each time $a$ crosses a saddle, the basin with a large birth merges with that with a small birth.
In the fourth row in Fig.~\ref{fig:filtration_PES}, basins ``$0$'' and ``$2$'' are merged.
In this case, basin ``$2$'' has a larger birth $b_2$ than the basin ``$0$'' $b_0$. Therefore, we consider that basin ``$2$'' merged into basin ``$0$'', and basin ``$0$'' survived.
Thus, the death of basin ``$2$'' $d_2$ has been introduced. 
In the same way, the death of basins, $d_1$ will be introduced (fifth row in Fig.~\ref{fig:filtration_PES}).
The basin ``$0$'' survives forever; thus, we define $d_0=\infty$.
The collection $\{(b_i,d_i)\}$ is a so-called persistence diagram. 
The persistence diagram of the basin is especially called the 0-th persistence diagram.

The merged basins sometimes form a closed path with impassable space inside (sixth row in Fig.~\ref{fig:filtration_PES}). 
Such a closed path is another object of homology.
The sixth row in Fig.~\ref{fig:filtration_PES} describes the energy where the path labeled ``$3$'' appears.
We define energy $a$ as the birth of path $b_3$.
Finally, the impassable space vanishes when $3$ crosses a second-order saddle point (the last row in Fig.~\ref{fig:filtration_PES}).
We define energy $a$ as the death of path $d_3$.
Many closed paths can exist in a more complicated PES than the example shown in Fig.~\ref{fig:filtration_PES}.
The collection $\{(b_i,d_i)\}$ for closed paths is called the 1-st persistence diagram.

However, the description of chemical reactions does not include higher-order PES saddle points within the transition state theory.
The GRRM program extracts only the zeroth and first-order saddles.
Thus, death in the 1-st persistence diagram cannot be determined. 
To apply PH to RRM, topological features associated with higher-order saddles should properly be discarded.
In other words, it is necessary to develop a new method that targets only RRM and not PES.
In the next section, we present a theoretical treatment to construct a PH applicable to RRM data.
Here, we introduce a parameter $\varepsilon$ to preserve information on detours (e.g., stepwise and concerted pathways) expressed in 1-st persistence diagrams.
This is because these pathways, e.g., contribute entropically to the reaction rate.

\subsection{Rigorous Mathematical Method}
\label{sec:math}
In this section, we review the theoretical background of PH for the filtered clique complex of a weighted undirected simple graph, which is a mathematical expression of a simplified RRM.
PH is a mathematical tool that extracts topological features from discrete data. 
In our application such input data are weighted graphs. 
We constructed a topological space from a weighted graph called a \textit{clique complex}.
This is the set of \textit{cliques} and allows us to extract higher-dimensional information from a graph. 
A nested structure called a \textit{filtration} is required to construct a PH. 
The clique complex can be equipped with a filtration determined by a weighting of the graph. 
The information obtained by PH can vary depending on the way the filtration is designed.
Our filtration is adjusted using a small constant $\varepsilon$ to capture when cliques appear, although there is a natural filtration ($\varepsilon = 0$ case) by a weighting.
PH is reviewed in Refs.~\citenum{edelsbrunner2010computational} and \citenum{otter2017roadmap}.

\subsubsection{Simple graphs and clique complexes}
Before explaining the clique complex, we introduce the terminology of graph theory \cite{Diestel2017_GraphTheory, Kozlov2008_CombinatorialAlgebraicTopology}. 
A (undirected) simple graph $G = (V, E)$ is a pair of sets: a set of \textit{vertices}, $V$, and a set of \textit{edges}, $E \subset \{ \{u, v\} \subset V \mid  u \not=v \}$.
By definition, simple graphs do not contain loops or multiple edges. 
In an RRM, $V$ and $E$ correspond to the set of EQs and TSs, respectively.
We denote the edge with endpoints, $u$ and $v$, as $\overline{uv}$.
We assume that the vertex set $V$ is finite (therefore, the edge set $E$ is finite).
A pair $G' = (V', E')$ is a \textit{subgraph} of $G$ if $V' \subset V$ and $E' \subset \{\overline{uv} \in E \mid u, v \in V' \}$.
We write $G' \subset G$ if $G'$ is a subgraph of $G$.

Herein, we introduce the concepts of cliques and clique complexes.
For an integer $p \geq 2$, a $p$-\textit{clique} in a graph $G = (V, E)$ is a subset $S$ of $V$ such that $S$ contains $p$ vertices and each distinct pair $u, v \in S$ is connected by an edge in $E$.
A $1$-clique is a singleton set $\{v\}$ that contains vertex $v \in V$.
The collection of all cliques in $G$ is called the \textit{clique complex} of $G$, denoted by $X(G)$.

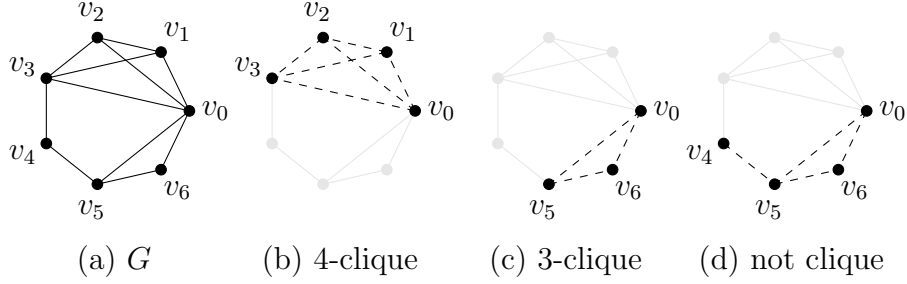
\begin{figure}[ht]
	\centering
    \def\n{6}
\def\edgelist{0/1,0/2,0/3,0/6,1/2,1/3,2/3,3/4,4/5,0/5,5/6}
\begin{tikzpicture}
	\foreach \t in {0,1,...,\n} {
		\coordinate (\t) at ({360*\t/(\n+1)}:1);
		\draw [fill=black] (\t) circle (2pt);
		\node at ({360*\t/(\n+1)}:1.35) {$v_{\t}$};
	}
    \foreach \u/\v in \edgelist {
        \draw (\u)--(\v);
    }
    \node at (0,-2) {(a) $G$};

\begin{scope}[xshift=3cm]
	\foreach \t in {0,1,...,\n} {
		\coordinate (\t) at ({360*\t/(\n+1)}:1);
	}
    \foreach \u/\v in {0/1,0/2,0/3,1/2,1/3,2/3} {
        \draw [dashed] (\u)--(\v);
    }
    \foreach \u/\v in {0/6,3/4,4/5,0/5,5/6} {
        \draw [black!10] (\u)--(\v);
    }
	\foreach \t in {0,1,...,3} {
		\draw [fill=black] (\t) circle (2pt);
        \node at ({360*\t/(\n+1)}:1.35) {$v_{\t}$};
	}
	\foreach \t in {4,...,\n} {
		\filldraw [black!10] (\t) circle (2pt);
	}
    \node at (0,-2) {(b) $4$-clique};
\end{scope}
\begin{scope}[xshift=6cm]
	\foreach \t in {0,1,...,\n} {
		\coordinate (\t) at ({360*\t/(\n+1)}:1);
	}
    \foreach \u/\v in {3/4,0/1,0/2,0/3,1/2,1/3,2/3,4/5} {
        \draw [black!10] (\u)--(\v);
    }
    \foreach \u/\v in {0/6,0/5,5/6} {
        \draw [dashed] (\u)--(\v);
    }
	\foreach \t in {0,5,6} {
		\draw [fill=black] (\t) circle (2pt);
        \node at ({360*\t/(\n+1)}:1.35) {$v_{\t}$};
	}
	\foreach \t in {1,2,3,4} {
		\filldraw [black!10] (\t) circle (2pt);
	}
    \node at (0,-2) {(c) $3$-clique};
\end{scope}

\begin{scope}[xshift=9cm]
	\foreach \t in {0,1,...,\n} {
		\coordinate (\t) at ({360*\t/(\n+1)}:1);
	}
    \foreach \u/\v in {3/4,0/1,0/2,0/3,1/2,1/3,2/3} {
        \draw [black!10] (\u)--(\v);
    }
    \foreach \u/\v in {0/6,4/5,0/5,5/6} {
        \draw [dashed] (\u)--(\v);
    }
	\foreach \t in {0,4,5,6} {
		\draw [fill=black] (\t) circle (2pt);
        \node at ({360*\t/(\n+1)}:1.35) {$v_{\t}$};
	}
	\foreach \t in {1,2,3} {
		\filldraw [black!10] (\t) circle (2pt);
	}
    \node at (0,-2) {(d) not clique};
\end{scope}

\end{tikzpicture}
	\caption{(a) An example of graph $G$; in (b), the set of vertices $\{v_0, v_1, v_2, v_3\}$ forms a $4$-clique in the graph $G$ since each two vertices are connected by an edge (indicated by a dashed line). Similarly, (c) is a $3$-clique. However, (d) is not a clique since $v_0$ and $v_4$ are not connected.}
    \label{fig:cliques}
\end{figure}

For example, consider the graph $G=(V,E)$ in Fig.~\ref{fig:cliques} (a). 
The set $\{v_0, v_1, v_2, v_3\}$ in Fig.~\ref{fig:cliques} (b) is a $4$-clique, and $\{v_0, v_5, v_6\}$ in Fig.~\ref{fig:cliques} (c) is a $3$-clique because each distinct pair of vertices is connected by an edge (represented by a dashed line).
Note that all non-empty subsets of a clique are cliques; e.g., $\{v_0, v_1, v_2\}$ is a $3$-clique. 
In addition, each singleton set $\{v\}$ of vertex $v \in V$ is a $1$-clique, and each edge in $E$ is a $2$-clique.
However, $\{v_0, v_4, v_5, v_6\}$ in Fig.~\ref{fig:cliques} (d) is not a clique since, for instance, $v_0$ and $v_4$ are not directly connected.

\subsubsection{Weighting and filtration}
To express the energy values in graphs and clique complexes, we introduce \textit{weighting} on a graph, which is defined as a pair of real-valued functions $w_V \colon V \to \mathbf{R}$ and $w_E \colon E \to \mathbf{R}$.
We denote both weight functions by $w$ if it is clear from the context whether $w$ refers to $w_V$ or $w_E$.
For our purposes, we consider only the weight function $w \colon G \to \mathbf{R}$ that satisfies the condition
\begin{equation}\label{weight_requirement}
	w_V (u), w_V (v) \leq w_E (\overline{uv}) \text{ for } \overline{uv} \in E.
\end{equation}
Subsequently, we define a weighting $w^G \colon X(G) \to \mathbf{R}$ on the clique complex as follows: 
for a $p$-clique $\sigma = \{v_{i_1}, \dots, v_{i_p}\}$, ($i_1 < \dots < i_p$), 
\begin{equation}\label{eq:clique-weight}
    w^G (\sigma; \varepsilon) = 
    \begin{cases}
        w_V (v_{i_1}) & (p = 1) \\
        w_E (\overline{v_{i_1} v_{i_2}}) & (p = 2) \\
        \max \{ w^G (\tau; \varepsilon) \mid \tau \subsetneq \sigma \} + \varepsilon & (p \geq 3), 
    \end{cases}
\end{equation}
or equivalently,
\begin{equation}\label{eq:clique-weight2}
    w^G (\sigma; \varepsilon) = 
    \begin{cases}
        w_V (v_{i_1}) & (p = 1) \\
        \max \{ w_E (\overline{v_{i_j} v_{i_k}}) \mid 1 \leq j < k \leq p \} + (p - 2) \varepsilon & (p \geq 2),
    \end{cases}
\end{equation}
where $\varepsilon$ denotes a parameter with a non-negative real number.
Henceforce, $w^G (\sigma; \varepsilon)$ is simply referred to as $w^G (\sigma)$.
By construction, the weighting $w^G$ satisfies the following condition:
\begin{equation}\label{eq:weightcondition}
    w^G (\tau) \leq w^G (\sigma)
    \text{ for cliques }
    \tau \subset \sigma.
\end{equation}

In this setting, we can construct an increasing sequence corresponding to the clique complex.
First, we define
\begin{equation}\label{eq:XofGofa-def}
    X(G)(a) := \{
        \sigma \in X(G)
    \mid
        w^G (\sigma) \leq a
    \}
\end{equation}
for a real number $a \in \mathbf{R}$.
By the condition of Eq.~(\ref{eq:weightcondition}), this is a subset of cliques of $G$ each of whose weights is less than or equal to $a$.

By construction, if $a < a'$, $X(G)(a) \subset X(G)(a')$.
Thus, corresponding to an increasing sequence $a_0 < a_1 < \dots < a_l$ of values of $w^G$, we obtain an increasing sequence
\begin{equation}\label{eq:XofG_sequence}
    X(G)(a_0) \subset X(G)(a_1) \subset \dots \subset X(G)(a_l).
\end{equation}
This sequence is called ($\varepsilon$-)\textit{adjusted weight rank clique filtration} on the clique complex $X(G)$ of a weighted graph $(G, w)$ with respect to $\varepsilon \geq 0$.
For $\varepsilon = 0$, this filtration coincides with the weight rank clique filtration proposed by Petri \textit{et al.} \cite{petri2013topological}, which treats graphs with real-value weights on the edges and all vertices appearing from the beginning. 
As explained in Sec.~\ref{Sec:toy-model}, the introduction of non-zero $\varepsilon$ can enrich the information derived from the PH analysis.

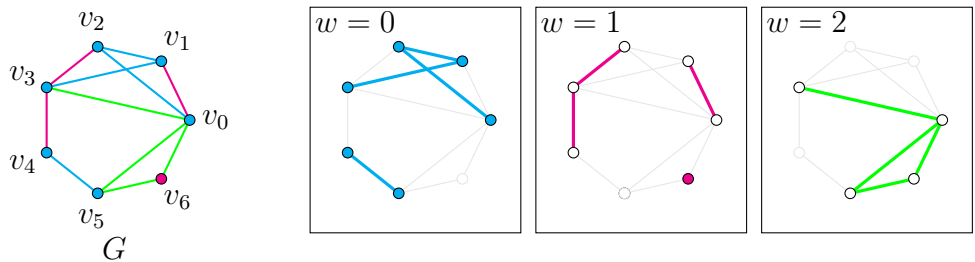
\begin{figure}[ht]
    \centering
    \def\labely{1.7} 
\def\nodes{0,1,2,3,4,5,6}
\def\n{7} 
\def\nodesa{0,1,2,3,4,5} 
\def\nodesb{6} 
\def\edges{0/2,1/2,1/3,4/5,0/1,2/3,3/4,0/3,0/5,0/6,5/6}
\def\edgesa{0/2,1/2,1/3,4/5} 
\def\edgesb{0/1,2/3,3/4} 
\def\edgesc{0/3,0/5,0/6,5/6} 

\begin{tikzpicture}[auto]
\node at (270:\labely) {$G$};
\foreach \t in \nodes {
	\coordinate (\t) at ({360*\t/(\n)}:1);
}

\foreach \u/\v in \edgesa {
    \draw [cyan, thick] (\u)--(\v);
}
\foreach \u/\v in \edgesb {
    \draw [magenta, thick] (\u)--(\v);
}
\foreach \u/\v in \edgesc {
    \draw [green, thick] (\u)--(\v);
}

\foreach \t in \nodesa {
	\draw [fill=cyan] (\t) circle (2pt);
	\node at ({360*\t/(\n)}:1.35) {$v_{\t}$};
}
\foreach \t in \nodesb {
	\draw [fill=magenta] (\t) circle (2pt);
	\node at ({360*\t/(\n)}:1.35) {$v_{\t}$};
}

\begin{scope}[xshift=1cm]
\begin{scope}[xshift=3cm]
\draw [thin] (-1.4,-1.5) rectangle (1.4,1.5);
\node at (-0.8,1.3) {$w = 0$};
\foreach \t in \nodes {
	\coordinate (\t) at ({360*\t/(\n)}:1);
}
\foreach \u/\v in \edgesa {
    \draw [cyan,very thick] (\u)--(\v);
}
\foreach \u/\v in \edgesb {
    \draw [black!10] (\u)--(\v);
}

\foreach \u/\v in \edgesc {
    \draw [black!10] (\u)--(\v);
}
\foreach \t in \nodesa {
	\draw [fill=cyan] (\t) circle (2pt);
}
\foreach \t in \nodesb {
	\draw [black!10,fill=white] (\t) circle (2pt);
}
\end{scope}

\begin{scope}[xshift=6cm]
\draw [thin] (-1.4,-1.5) rectangle (1.4,1.5);
\node at (-0.8,1.3) {$w = 1$};
\foreach \t in \nodes {
	\coordinate (\t) at ({360*\t/(\n)}:1);
}
\foreach \u/\v in \edgesa {
    \draw [black!10] (\u)--(\v);
}
\foreach \u/\v in \edgesc {
    \draw [black!10] (\u)--(\v);
}
\foreach \u/\v in \edgesb {
    \draw [magenta, very thick] (\u)--(\v);
}
\foreach \t in \nodesa {
	\draw [thin,fill=white] (\t) circle (2pt);
}
	\draw [black!10,fill=white] (5) circle (2pt);
\foreach \t in \nodesb {
	\draw [fill=magenta] (\t) circle (2pt);
}
\end{scope}

\begin{scope}[xshift=9cm]
\draw [thin] (-1.4,-1.5) rectangle (1.4,1.5);
\node at (-0.8,1.3) {$w = 2$};
\foreach \t in \nodes {
	\coordinate (\t) at ({360*\t/(\n)}:1);
}
\foreach \u/\v in \edgesb {
    \draw [black!10] (\u)--(\v);
}

\foreach \u/\v in \edgesc {
    \draw [green, very thick] (\u)--(\v);
}
\foreach \u/\v in \edgesa {
    \draw [black!10] (\u)--(\v);
}
\foreach \t in {0,3,5,6} {
	\draw [thin,fill=white] (\t) circle (2pt);
}
\foreach \t in {1,2,4} {
	\draw [black!10,fill=white] (\t) circle (2pt);
}
\end{scope}
\end{scope}

\end{tikzpicture}
    \caption{An example of a weighting $w$ on $G$. Vertices $v_0, v_1, \dots, v_5$ and edges $\overline{v_0 v_2}$, $\overline{v_1 v_2}$, $\overline{v_1 v_3}$, $\overline{v_4 v_5}$ have a weight value $0$. Similarly a vertex $v_6$ and edges $\overline{v_0 v_1}, \overline{v_2 v_3}, \overline{v_3 v_4}$ have a weight $1$ (white vertices have already appeared and just indicate the endpoints). Remained edges have $2$.}
    \label{fig:weighting_on_graph}
\end{figure}

Figure \ref{fig:weighting_on_graph} depicts an example of a weighting $w$ on graph $G$ depicted in Fig.~\ref{fig:cliques}.
In the three figures on the right, the filled-in vertices and thickened edges have corresponding weights $w = 0, 1$, and $2$, whereas the gray edges and unfilled vertices have different weights.
This weighting satisfies the condition \eqref{weight_requirement}, i.e., the weight of an edge is equal to or greater than the weight of each of its endpoints.
The resulting clique complex of $G$ is a $3$-dimensional polyhedron, consisting of a solid tetrahedron ($3$-dimensional), triangle ($2$-dimensional), edge ($1$-dimensional), and vertex ($0$-dimensional).
The weighting of the clique complex $X(G)$ induced by $w$ is determined as follows:
Let $\varepsilon > 0$ be sufficiently small, i.e., at least $\varepsilon < 1$.
The $1$- and $2$-cliques have the same weights as the corresponding vertices and edges.
For $3$-cliques, the weights are as follows:
\begin{equation*}
    w^G (\sigma) = 
    \begin{cases}
        1 + \varepsilon & \text{if } \sigma = \{ v_0, v_1, v_2 \}, \{ v_1, v_2, v_3 \} \\
        2 + \varepsilon & \text{if } \sigma = \{ v_0, v_1, v_3 \}, \{ v_0, v_2, v_3 \}, \{ v_0, v_5, v_6 \}.
    \end{cases}
\end{equation*}
Because the maximal weight of the cliques contained in a $4$-clique $\{ v_0, v_1, v_2, v_3 \}$ is $2 + \varepsilon$, the weight of $\{ v_0, v_1, v_2, v_3 \}$ is $2 + 2\varepsilon \ (> 2 + \varepsilon)$.
Based on the weighting, $w^G$, the adjusted weight rank clique filtration on $X(G)$ is of the following form:
\begin{equation*}
    X(G)(0) \subset X(G)(1) \subset X(G)(1 + \varepsilon) \subset X(G)(2) \subset X(G)(2 + \varepsilon) \subset X(G)(2 + 2 \varepsilon) = X(G).
\end{equation*}
\begin{figure}[h]
    \centering
    \def\labely{1.7} 
\def\nodes{0,1,2,3,4,5,6}
\def\n{7} 
\def\nodesa{0,1,2,3,4,5} 
\def\nodesb{6} 
\def\edges{0/2,1/2,1/3,4/5,0/1,2/3,3/4,0/3,0/5,0/6,5/6}
\def\edgesa{0/2,1/2,1/3,4/5} 
\def\edgesb{0/1,2/3,3/4} 
\def\edgesc{0/3,0/5,0/6,5/6} 
\def\edgess{1/2,1/3,4/5,0/1,2/3,3/4,0/3,0/5,0/6,5/6}
\def\edgest{1/2,1/3,4/5,0/1,2/3,3/4}

\begin{tikzpicture}[auto]
\node at (270:\labely) {$X(G)(0)$};
\foreach \t in \nodesa {
	\coordinate (\t) at ({360*\t/(\n)}:1);
}
    \draw [] (0)--($(0)!0.65!(2)$);
    \draw [] ($(0)!0.75!(2)$)--(2);
\foreach \u/\v in {1/2,1/3,4/5} {
    \draw (\u)--(\v);
}
\foreach \t in \nodesa {
	\draw [fill=black] (\t) circle (2pt);
	\node at ({360*\t/(\n)}:1.35) {$v_{\t}$};
}

\begin{scope}[xshift=4cm]
\node at (270:\labely) {$X(G)(1)$};

\foreach \t in \nodes {
	\coordinate (\t) at ({360*\t/(\n)}:1);
}
    \draw [] (0)--($(0)!0.65!(2)$);
    \draw [] ($(0)!0.75!(2)$)--(2);
\foreach \u/\v in {1/2,1/3,4/5} {
    \draw [] (\u)--(\v);
}
\foreach \u/\v in \edgesb {
    \draw [] (\u)--(\v);
}

\foreach \t in \nodesa {
	\draw [fill=black] (\t) circle (2pt);
}
\foreach \t in \nodesb {
	\draw [fill=black] (\t) circle (2pt);
	\node at ({360*\t/(\n)}:1.35) {$v_{\t}$};
}
\end{scope}

\begin{scope}[xshift=8cm]
\node at (270:\labely) {$X(G)(1 + \varepsilon)$};

\foreach \t in \nodes {
	\coordinate (\t) at ({360*\t/(\n)}:1);
}
\filldraw [gray, opacity=0.2] (0)--(1)--(2)--cycle;
    \draw [] (0)--($(0)!0.6!(2)$);
    \draw [dashed] ($(0)!0.6!(2)$)--(2);
\filldraw [gray, opacity=0.2] (1)--(2)--(3)--cycle;
\foreach \u/\v in \edgest {
    \draw [] (\u)--(\v);
}

\foreach \t in \nodes {
	\draw [fill=black] (\t) circle (2pt);
}
\end{scope}

\begin{scope}[yshift=-4cm]
\node at (270:\labely) {$X(G)(2)$};

\foreach \t in \nodes {
	\coordinate (\t) at ({360*\t/(\n)}:1);
}
\filldraw [gray, opacity=0.2] (0)--(1)--(2)--cycle;
    \draw [] (0)--($(0)!0.6!(2)$);
    \draw [dashed] ($(0)!0.6!(2)$)--(2);
\filldraw [gray, opacity=0.2] (1)--(2)--(3)--cycle;
\foreach \u/\v in \edgess {
    \draw [] (\u)--(\v);
}

\foreach \t in \nodes {
	\draw [fill=black] (\t) circle (2pt);
}
\end{scope}

\begin{scope}[xshift=4cm,yshift=-4cm]
\node at (270:\labely) {$X(G)(2 + \varepsilon)$};

\foreach \t in \nodes {
	\coordinate (\t) at ({360*\t/(\n)}:1);
}
    \draw [dashed] (0)--(2);
    \filldraw [gray, opacity=0.3] (0)--(1)--(3)--cycle;
    \filldraw [gray, opacity=0.2] (0)--(5)--(6)--cycle;
    \filldraw [gray, opacity=0.2] (1)--(2)--(3)--cycle;
\foreach \u/\v in \edgess {
    \draw [] (\u)--(\v);
}
\foreach \t in \nodes {
	\draw [fill=black] (\t) circle (2pt);
}
\end{scope}

\begin{scope}[xshift=8cm,yshift=-4cm]
\node at (270:\labely) {$X(G)(2 + 2 \varepsilon) = X(G)$};

\foreach \t in \nodes {
	\coordinate (\t) at ({360*\t/(\n)}:1);
}
    \draw [dotted] (0)--(2);
    \filldraw [gray, opacity=0.4] (0)--(1)--(3)--cycle;
    \filldraw [gray, opacity=0.2] (0)--(5)--(6)--cycle;
    \filldraw [gray, opacity=0.3] (1)--(2)--(3)--cycle;
\foreach \u/\v in \edgess {
    \draw [] (\u)--(\v);
}
\foreach \t in \nodes {
	\draw [fill=black] (\t) circle (2pt);
}
\end{scope}
\end{tikzpicture}
    \caption{Increasing sequence of $X(G)(a)$ defined in Eq.~(\ref{eq:XofGofa-def}) in the adjusted weight rank clique filtration on $X(G)$ determined by the weighting presented in Fig.~\ref{fig:weighting_on_graph} when the positive $\varepsilon$ parameter is adopted. This is analogous to the sequence of the topological representations (simplicial complexes) in Fig.~\ref{fig:filtration_PES}.}

    \label{fig:adjusted_wrc_filtration}
\end{figure}
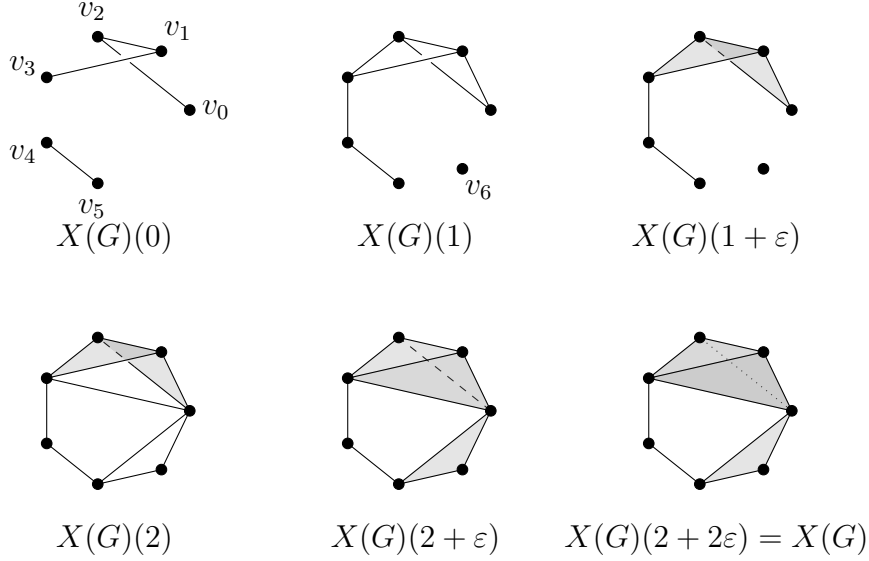
In Fig.~\ref{fig:adjusted_wrc_filtration}, $X(G)(0)$ comprises two connected components.
Vertex $v_6$ and edges $\overline{v_0 v_1}$, $\overline{v_2 v_3}$, and $\overline{v_3 v_4}$ appear in $X(G)(1)$.
Note that, e.g., 2-cliques $\{v_0, v_1\}$, $\{v_0, v_2\}$ and $\{v_1, v_2\}$ are contained in $X(G)(1)$, while the 3-clique $\{ v_0, v_1, v_2 \}$ is not since it has a weight greater than $1$. 
The shaded triangles following $X(G)(1 + \varepsilon)$ indicate the corresponding $3$-cliques that consist of the vertices of the triangle.
$X(G)(2 + \varepsilon)$ does not contain a filling (a $4$-clique) in the tetrahedron $v_0 v_1 v_2 v_3$ whereas $X(G)(2 + 2 \varepsilon)$ does.
Note that $X(G)(1)$ cannot be presented as a clique complex of certain graphs because though the existence of the edges ($2$-cliques) $\overline{v_0 v_1}$, $\overline{v_0 v_2}$, and $\overline{v_1 v_2}$ implies the existence of the $3$-clique $\{ v_0, v_1, v_2 \}$, $X(G)(1)$ does not contain $\{ v_0, v_1, v_2 \}$.
Similarly, $X(G)(2)$ and $X(G)(2 + \varepsilon)$ cannot be realized as clique complexes.
In general, $X(G)(a)$ in the adjusted weight rank clique filtration is not necessarily a clique complex, whereas the weight rank clique filtration ($\varepsilon = 0$) consists of the clique complexes of subgraphs of $G$. 

In particular, the weight rank clique filtration ($\varepsilon = 0$) can be defined by the sequence of subgraphs of graph $G$.
For $a \in \mathbf{R}$, we define
\begin{equation}\label{eq:Gofa-definition}
    V(a) = \{ v \in V \mid w (v) \leq a \}
    \quad \text{and} \quad
    E(a) = \{ e \in E \mid w (e) \leq a \}.
\end{equation}
By Eq.~(\ref{eq:weightcondition}), $G(a) = (V(a), E(a))$ forms a graph.
If $a < a'$, then $G(a) \subset G(a')$.
For a weighted graph $(G, w)$ and an increasing sequence $a_0 < a_1 < \dots < a_l$, we have the following sequence of subgraphs:
\begin{equation*}
	G (a_0) \subset G (a_1) \subset \dots \subset G (a_l).
\end{equation*}
By considering the clique complex for each graph $G (a_k)$, we obtain the following sequence of clique complexes:
\begin{equation*}
	X (G(a_0)) \subset X (G(a_1)) \subset \dots \subset X (G(a_l)).
\end{equation*}
This sequence is equivalent to that in Eq.~(\ref{eq:XofG_sequence}) for $\varepsilon=0$.

\begin{figure}[h]
	\centering
	\def\n{6} 
\def\labely{1.7} 

\begin{tikzpicture}
\node at (270:\labely) {$G(0)$};

\foreach \t in {0,1,...,5} {
	\coordinate (\t) at ({360*\t/(\n+1)}:1);
	\draw [fill=black] (\t) circle (2pt);
	\node at ({360*\t/(\n+1)}:1.35) {$v_{\t}$};
}

\draw (0)--(2);
\draw (1)--(2);
\draw (1)--(3);
\draw (4)--(5);

\node at (2,0) {$\subset$};

\begin{scope}[xshift=4cm]
\node at (270:\labely) {$G(1)$};

\foreach \t in {0,1,...,\n} {
	\coordinate (\t) at ({360*\t/(\n+1)}:1);
	\draw [fill=black] (\t) circle (2pt);
}

\node at ({360*6/7}:1.35) {$v_6$};

\draw (0)--(1);
\draw (0)--(2);
\draw (1)--(2);
\draw (1)--(3);
\draw (2)--(3);
\draw (3)--(4);
\draw (4)--(5);
\end{scope}

\node at (6,0) {$\subset$};

\begin{scope}[xshift=8cm]
\node at (270:\labely) {$G(2)$};

\foreach \t in {0,1,...,\n} {
	\coordinate (\t) at ({360*\t/(\n+1)}:1);
	\draw [fill=black] (\t) circle (2pt);
}

\draw (0)--(1);
\draw (0)--(2);
\draw (0)--(3);
\draw (0)--(6);
\draw (1)--(2);
\draw (1)--(3);
\draw (2)--(3);
\draw (3)--(4);
\draw (4)--(5);
\draw (0)--(5);
\draw (5)--(6);
\end{scope}
\end{tikzpicture}
	\caption{Increasing sequence of subgraphs $G(a)=(V(a), E(a))$ defined by Eq.~(\ref{eq:Gofa-definition}) determined by the weighting presented in Fig.~\ref{fig:weighting_on_graph}.}
	\label{subgraph-sequence}
\end{figure}
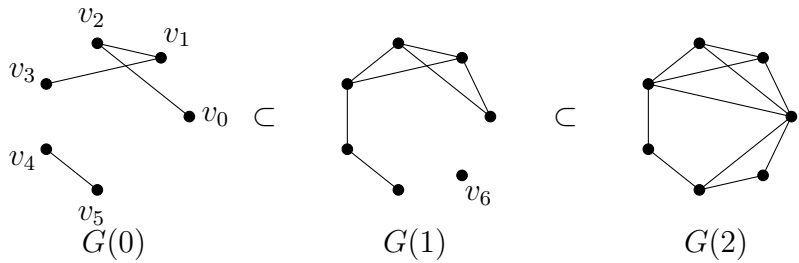

For example, we have a sequence of subgraphs above $G$, as shown in Fig.~\ref{subgraph-sequence}, using the weighting defined in Fig.~\ref{fig:weighting_on_graph}.
The adjusted weight rank clique filtration on $X(G)$ is shown in Fig.~\ref{fig:adjusted_wrc_filtration} and the weight rank clique filtration corresponds to the case, $\varepsilon = 0$, which consists of $X(G(0)) = X(G)(0)$, $X(G(1)) = X(G)(1) \cup X(G)(1 + \varepsilon)$, and $X(G(2)) = X(G)(2) \cup X(G)(2 + \varepsilon) \cup X(G)(2 + 2\varepsilon)$.

\subsubsection{Persistent homology and its visualization}
A clique complex has a \textit{simplicial complex} structure.
Given a filtration on a simplicial complex, PH can be defined in general.
Here, we briefly summarize the terminology used in PH theory.
Further details of this theory can be found in Refs.~\citenum{edelsbrunner2010computational} and \citenum{otter2017roadmap}.

We denote the $p$-th homology group of a clique complex $X$ with coefficients in $\mathbf{Z} / 2 \mathbf{Z} = \{0, 1\}$ ($\mathbf{Z}$ is the ring of integers) \cite{Wallace-book} by $H_p (X)$.
For a filtered clique complex $X$ with filtration 
\begin{equation*}
     X(0) \subset X (1) \subset \cdots \subset X(l), 
\end{equation*}
we denote the homomorphism induced by the inclusion $X(i) \subset X(j)$ by $\iota_i^j \colon H_\ast (X(i)) \to H_\ast (X(j))$.
The $p$-\textit{th persistent homology} of the filtered clique complex $X$ is the pair, $\{ H_p (X(i)) \}_i$ and $\{ \iota_i^j \}_{i \leq j}$.
An element $\alpha \in H_p (X(i))$ is \textit{born} in $X(i)$ ($\alpha \not= 0$) if there is no $\beta \in H_p (X(i - 1))$ such that $\iota_{i-1}^i (\beta) = \alpha$, and it \textit{dies} in $X(j)$ ($i < j$) if $j$ is the smallest index for which $\iota_i^j (\alpha) = 0$.
It is always possible to define the set of generators of $H_p (X(i))$ following the \textit{elder rule} (like as the first-in-last-out stack, the element born last dies first, i.e., the elements born earlier (\textit{elder} elements) survive), which is simply referred to as the set of \textit{generators} in this paper.
If $\alpha$ is born in $X(i)$ but never dies, we say that $\alpha$ \textit{lives forever}.

Each generator $\alpha$ of $H_p (X(i))$ may be associated with the point $(a_i, a_j)$ in the Euclidean plane $\mathbf{R}^2$ if $\alpha$ is born in $H_p (X(i))$ and dies in $H_p (X(j))$.
This information can be summarized using persistence diagrams and persistence barcodes.
Henceforth, we refer to $b_\alpha = a_i$ as the \textit{birth} and to $d_\alpha = a_j$ as the \textit{death} of the generator $\alpha$. 
If $\alpha$ is born in $H_p (X(i))$ and lives forever, we associate $\alpha$ with the point $(a_i, \infty) \in (\mathbf{R} \cup \{\infty\})^2$.
The $p$-\textit{th persistence diagram} of $X$, $\mathcal{D}_p (X)$, is the multiset of points $(a_i, a_j) \in (\mathbf{R} \cup \{\infty\})^2$ corresponding to generators that are born in $H_p (X(i))$ and either die in $H_p (X(j))$ or live forever.
The $p$-\textit{th persistence barcode} of $X$, $\mathcal{B}_p (X)$, is the multiset of half-open intervals $[a_i, a_j) \subset \mathbf{R} \cup \{\infty\}$ corresponding to generators that are born in $H_p (X(i))$ and either die in $H_p (X(j))$ or live forever.

\begin{figure}[H]
	\centering
	\begin{tikzpicture}
\draw [->] (0,0)node[below left]{$0$}--(3,0);
\draw [->] (0,0)--(0,3);
\draw (0,0)--(2.8,2.8);

\draw [fill=black] (0,1)node[left]{$a_1$} circle (2pt);
\draw [fill=black] (1,2) circle (2pt);
\draw [] (0,3.2)node[left]{$\infty$} circle (2pt);

\draw [dotted] (0,2)node[left]{$a_2$} -- (1,2) -- (1,0)node[below]{$a_1$};

\node at (1.5,-1) {$\mathcal{D}_0 (X)$};

\begin{scope}[yshift=-2cm]
\draw [->] (0,0)--(3,0);
    \draw [thin] (0,-0.1)--(0,0.1)node[above]{$0$};
    \draw [thin] (1,-0.1)--(1,0.1)node[above]{$a_1$};
    \draw [thin] (2,-0.1)--(2,0.1)node[above]{$a_2$};
\draw [very thick] (0,-0.5)--(3,-0.5);
\draw [very thick] (0,-1)--(1,-1);
\draw [very thick] (1,-1.5)--(2,-1.5);

\node at (1.5,-2) {$\mathcal{B}_0 (X)$};
\end{scope}

\begin{scope}[xshift=5cm]
\draw [->] (0,0)node[below left]{$0$}--(3,0);
\draw [->] (0,0)--(0,3);
\draw (0,0)--(2.8,2.8);

\draw [] (2,3.2) circle (2pt);
\draw [dotted] (0,3.2)node[left]{$\infty$} -- (2,3.2) -- (2,0)node[below]{$a_2$};

\node at (1.5,-1) {$\mathcal{D}_1 (X)$};
\end{scope}

\begin{scope}[xshift=5cm,yshift=-2cm]
\draw [->] (0,0)--(3,0);
    \draw [thin] (0,-0.1)--(0,0.1)node[above]{$0$};
    \draw [thin] (2,-0.1)--(2,0.1)node[above]{$a_2$};
\draw [very thick] (2,-0.5)--(3,-0.5);

\node at (1.5,-2) {$\mathcal{B}_1 (X)$};
\end{scope}

\draw [dashed] (9.5,2.5)--(9.5,-3.5);

\begin{scope}[xshift=11cm]
\draw [->] (0,0)node[below left]{$0$}--(3,0);
\draw [->] (0,0)--(0,3);
\draw (0,0)--(2.8,2.8);

\node at (0,1)[left]{$a_1$};
\draw [dotted] (0,1)--(1,1);
\draw [fill=black,opacity=0.3] (0,1) circle (2pt);
\draw [fill=black] (1,2) circle (2pt);
\draw [opacity=0.3] (0,3.2)node[left]{$\infty$} circle (2pt);

\draw (1.15,2) to[bend left=30] (1.15,1);

\draw [dotted] (0,2)node[left]{$a_2$} -- (1,2) -- (1,0)node[below]{$a_1$};

\node at (1.5,-1) {$\mathcal{D}_0 (X)$};

\node (L) at (4,-1) [align=center]{length\\$a_2 - a_1$};

\draw [->] (L) to[bend right=30] (1.4,1.5);
\draw [->] (L) to[bend left=30] (2.1,-3.5);

\begin{scope}[yshift=-2cm]
\draw [->] (0,0)--(3,0);
    \draw [thin] (0,-0.1)--(0,0.1)node[above]{$0$};
    \draw [thin] (1,-0.1)--(1,0.1)node[above]{$a_1$};
    \draw [thin] (2,-0.1)--(2,0.1)node[above]{$a_2$};
\draw [very thick,opacity=0.3] (0,-0.5)--(3,-0.5);
\draw [very thick,opacity=0.3] (0,-1)--(1,-1);
\draw [dotted] (1,0)--(1,-1.5);
\draw [dotted] (2,0)--(2,-1.5);
\draw [very thick] (1,-1.5)--(2,-1.5);

\node at (1.5,-2) {$\mathcal{B}_0 (X)$};

\end{scope}
\end{scope}

\end{tikzpicture}
	\caption{$0$-th and $1$-st persistence diagrams and barcodes of the clique complex $X = X(G)$ constructed from a weighted graph $(G, w)$ in Fig.~\ref{fig:weighting_on_graph}}
	\label{PDsample}
\end{figure}
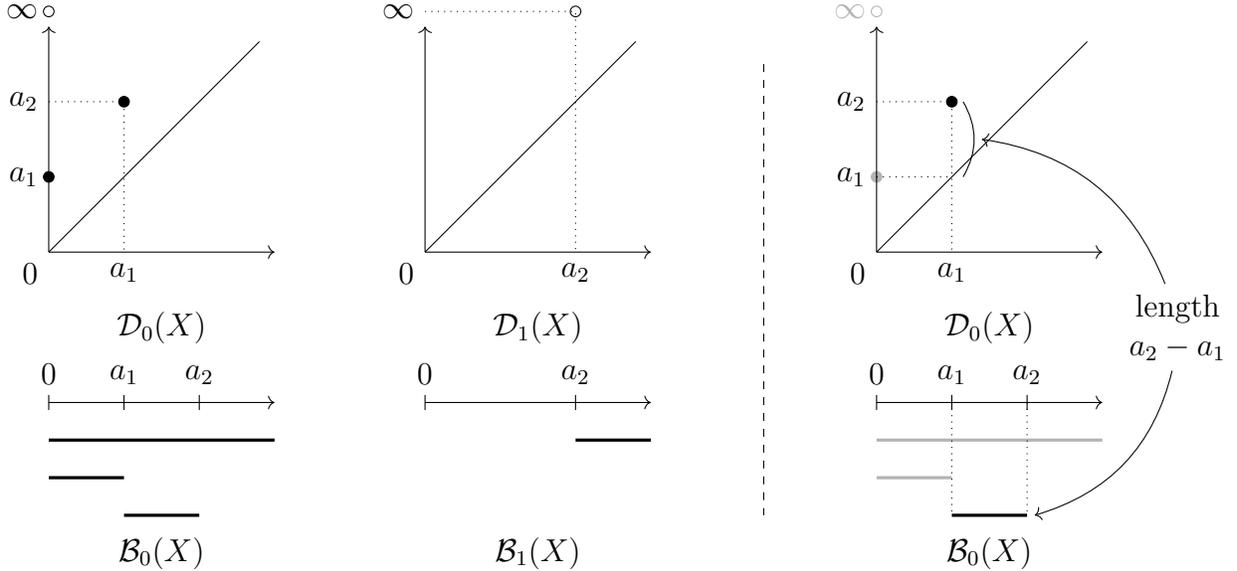

In the case of the filtered clique complex constructed from the sequence of subgraphs in Fig.~\ref{subgraph-sequence}, 
$0$-dimensional homology classes correspond to the connected components in $G(0)$. 
These are born in $H_0 (X(0))$, two of them die in $H_0 (X(1))$, and the class corresponding to $v_6$ is born in $H_0 (X(1))$ and dies in $H_0 (X(2))$.
The $1$-dimensional homology class corresponding to rectangle $v_0 v_3 v_4 v_5$ is born in $H_1 (X(2))$ and lives forever.
Persistence diagrams and barcodes were obtained for this example (Fig.~\ref{PDsample}).
The length of the bar corresponding to point $(a_i, a_j)$ in the persistence diagram is $a_j - a_i$. 
This coincides with the length of the line segment between $(a_i, a_j)$ and the intersection of the diagonal and vertical lines from $(a_i, a_j)$.
Here, the elder rule is used to draw persistence diagrams. 
According to this rule, when two components are merged, the previously born component is retained.

\begin{figure}
    \centering
    \begin{tikzpicture}
\draw [->] (0,0)node[below left]{$0$}--(3,0);
\draw [->] (0,0)--(0,3);
\draw (0,0)--(2.8,2.8);

\draw [fill=black] (0,1)node[left]{$a_1$} circle (2pt);
\draw [fill=black] (1,2) circle (2pt);
\draw [] (0,3.2)node[left]{$\infty$} circle (2pt);

\draw [dotted] (0,2)node[left]{$a_2$} -- (1,2) -- (1,0)node[below]{$a_1$};

\node at (1.5,4) {$\mathcal{D}_0 (X)$};

\begin{scope}[xshift=4.5cm]
\draw [->] (0,0)node[below left]{$0$}--(3,0);
\draw [->] (0,0)--(0,3);
\draw (0,0)--(2.8,2.8);

\draw [] (2,3.2) circle (2pt);
\draw [dotted] (0,3.2)node[left]{$\infty$} -- (2,3.2) -- (2,0)node[below]{$a_2$};

\node at (1.5,4) {$\mathcal{D}_1 (X)$};
\end{scope}

\draw [->] (8,1.5)--node[above]{put together}(10,1.5);

\begin{scope}[xshift=11cm]
\draw [->] (0,0)node[below left]{$0$}--(3,0);
\draw [->] (0,0)--(0,3);
\draw (0,0)--(2.8,2.8);

\node at (1.5,4) {$\mathcal{D}_0 (X)$};

\node at (0,1)[left]{$a_1$};
\draw [dotted] (0,1)--(1,1);
\draw [fill=black] (0,1) circle (2pt);
\draw [fill=black] (1,2) circle (2pt);
\draw [] (0,3.2)node[left]{$\infty$} circle (2pt);

\draw [dotted] (1,2) -- (1,0)node[below]{$a_1$};

\node at (4,1.5) {$\mathcal{D}_1 (X)$};

\draw [] (3.2,2) circle (2pt);
\draw [dotted] (3.2,0)node[below]{$\infty$} -- (3.2,2) -- (0,2)node[left]{$a_2$};

\end{scope}

\end{tikzpicture}
    \caption{Concise representation of the $0$-th and $1$-st PD in Fig.~\ref{PDsample}.}
    \label{fig:pd-presentation}
\end{figure}
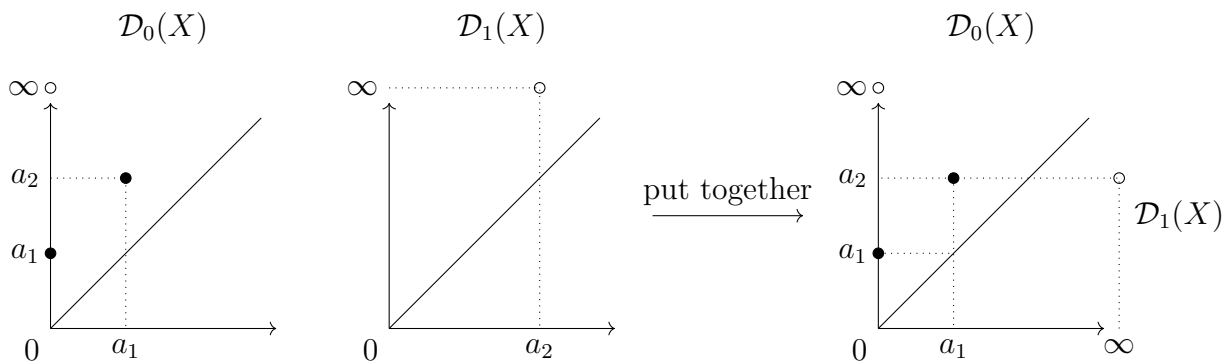

In the following argument, we focus on the $0$-th and $1$-st PH and their diagrams.
To make the figures concise, we depict the $0$-th and $1$-st persistence diagrams together---the upper half region presents the $0$-th persistence diagram (horizontal axis represents birth and vertical axis represents death) and the lower half region presents the $1$-st persistence diagram (vertical axis represents birth and horizontal axis represents death) in Fig.~\ref{fig:pd-presentation}.

\subsection{Approximation of the entire RRM graph using a simple graph}

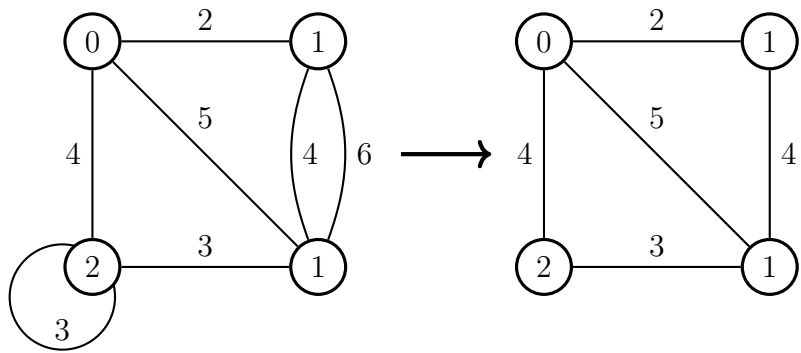
\begin{figure}
    \centering
    \begin{tikzpicture}

\draw[thick] (-1.4,-.4) circle (.7cm) node[below=0.15cm]{3};

\node[draw,shape=circle, very thick] (A) at (-1,3) {0};
\node[draw,fill=white,shape=circle, very thick] (B) at (-1,0) {2};
\node[draw,shape=circle, very thick] (C) at (2,0) {1};
\node[draw,shape=circle, very thick] (D) at (2,3) {1};

\draw[thick] (A) to node[left]{4} (B);
\draw[thick] (A) to node[above=0.2cm]{5} (C);
\draw[thick] (A) to node[above]{2} (D);
\draw[thick] (B) to node[above]{3} (C);
\draw[thick] (C) to[bend left =20] node[right]{4} (D);
\draw[thick] (C) to[bend right=20] node[right]{6} (D);

\draw[->,ultra thick](
3.1,1.5)--(4.3,1.5);

\node[draw, shape=circle, very thick] (a) at (5,3) {0};
\node[draw, shape=circle, very thick] (b) at (5,0) {2};
\node[draw, shape=circle, very thick] (c) at (8,0) {1};
\node[draw, shape=circle, very thick] (d) at (8,3) {1};

\draw[thick] (a) to node[left]{4} (b);
\draw[thick] (a) to node[above=0.2cm]{5} (c);
\draw[thick] (a) to node[above]{2} (d);
\draw[thick] (b) to node[above]{3} (c);
\draw[thick] (c) to node[right]{4} (d);



\end{tikzpicture}
    \hspace{0.5cm}
    \caption{(left) RRM containing loops and multiple edges. The number associated with each vertex or edge expresses its energy (or equivalently, weight). (right) Its simplified RRM.}
    \label{fig:simple-graph}
\end{figure}

As the entire RRM graph, $G^\mathrm{entire}$ may include loops and/or multiple edges for most systems, the adjusted weight rank clique filtration cannot be applied to $G^\mathrm{entire}$ directly.
However, as an approximate treatment, the filtration can be applied to a simple subgraph $G$ associated with the entire RRM graph $G^\mathrm{entire}$.
This was realized by eliminating loop edges and multiple edges other than those with the lowest energy weights (Fig.~\ref{fig:simple-graph}).

When the weight on each vertex or edge represents the corresponding potential energies, the requirement of Eq.~(\ref{weight_requirement}) is satisfied because the potential energy at the TS is higher than that at EQs connected by the corresponding IRC.
Note that this requirement may not be satisfied when the corrected energy, such as the Gibbs free energy, is represented by the weight value.

\subsection{Demonstration of the proposed method on the RRM of a toy model \label{Sec:toy-model}}

\begin{figure}[htbp]
  \begin{minipage}[b]{0.31\linewidth}
    \begin{tikzpicture}

\node[draw, shape=circle, very thick] (a) at (0,3) {0} node at (0,3) [left=0.25cm]{$v_1$};
\node[draw, shape=circle, very thick] (b) at (0,0) {2} node at (0,0) [left=0.25cm]{$v_4$};
\node[draw, shape=circle, very thick] (c) at (3,0) {1} node at (3,0) [right=0.25cm]{$v_3$};
\node[draw, shape=circle, very thick] (d) at (3,3) {1} node at (3,3) [right=0.25cm]{$v_2$};

\draw[thick] (a) to node[left]{$\overline{v_1v_4}$} node[right]{4}(b);
\draw[thick] (a) to node[above=0.15cm]{$\overline{v_1v_3}$} node[below]{5} (c);
\draw[thick] (a) to node[above]{$\overline{v_1v_2}$} node[below]{2} (d);
\draw[thick] (b) to node[above]{$\overline{v_3v_4}$} node[below]{3} (c);
\draw[thick] (c) to node[left]{$\overline{v_2v_3}$} node[right]{4} (d);



\end{tikzpicture}
    \subcaption{}
  \end{minipage}\\
    \begin{minipage}[b]{\linewidth}
    \centering
    \begin{tikzpicture}[scale=0.6, transform shape]

\node[draw, shape=circle, very thick] (a) at (0,3) {0};

\draw[->,ultra thick](
3.5,1.5)--(4.5,1.5);

\node[draw, shape=circle, very thick] (A4) at (5,3) {0};
\node[draw, shape=circle, very thick] (C4) at (8,3) {1};
\node[draw, shape=circle, very thick] (D4) at (8,0) {1};

\draw[->,ultra thick](
8.5,1.5)--(9.5,1.5);

\node[draw, shape=circle, very thick] (A3) at (10,3) {0};
\node[draw, shape=circle, very thick] (B3) at (10,0) {2};
\node[draw, shape=circle, very thick] (C3) at (13,3) {1};
\node[draw, shape=circle, very thick] (D3) at (13,0) {1};

\draw[thick] (A3) to node[below]{2} (C3);

\draw[->,ultra thick](
11.5,-0.5)--(11.5,-1.5);

\node[draw, shape=circle, very thick] (A2) at (10,-2) {0};
\node[draw, shape=circle, very thick] (B2) at (10,-5) {2};
\node[draw, shape=circle, very thick] (C2) at (13,-5) {1};
\node[draw, shape=circle, very thick] (D2) at (13,-2) {1};

\draw[thick] (A2) to node[below]{2} (D2);
\draw[thick] (B2) to node[below]{3} (C2);

\draw[->,ultra thick](
9.5,-3.5)--(8.5,-3.5);

\node[draw, shape=circle, very thick] (A1) at (5,-2) {0};
\node[draw, shape=circle, very thick] (B1) at (5,-5) {2};
\node[draw, shape=circle, very thick] (C1) at (8,-5) {1};
\node[draw, shape=circle, very thick] (D1) at (8,-2) {1};

\draw[thick] (A1) to node[right]{4} (B1);
\draw[thick] (A1) to node[below]{2} (D1);
\draw[thick] (B1) to node[below]{3} (C1);
\draw[thick] (C1) to node[right]{4} (D1);

\draw[->,ultra thick](
4.5,-3.5)--(3.5,-3.5);

\node[draw, shape=circle, very thick] (A) at (0,-2) {0};
\node[draw, shape=circle, very thick] (B) at (0,-5) {2};
\node[draw, shape=circle, very thick] (C) at (3,-5) {1};
\node[draw, shape=circle, very thick] (D) at (3,-2) {1};

\draw[thick] (A) to node[right]{4} (B);
\draw[thick] (A) to node[below]{5} (C);
\draw[thick] (A) to node[below]{2} (D);
\draw[thick] (B) to node[below]{3} (C);
\draw[thick] (C) to node[right]{4} (D);

\end{tikzpicture}
    \subcaption{}
  \end{minipage}\\
  \begin{minipage}[b]{0.4\linewidth}
    \centering
    \includegraphics[keepaspectratio, 
    scale=0.4
    ]{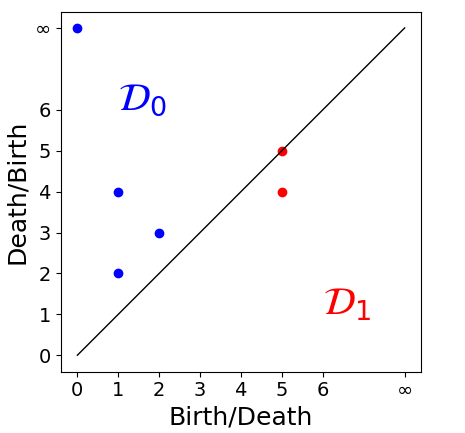}
    \subcaption{}
  \end{minipage}\\
  \begin{minipage}[b]{0.4\linewidth}
    \centering
    \includegraphics[keepaspectratio, 
    scale = 0.30
    ]{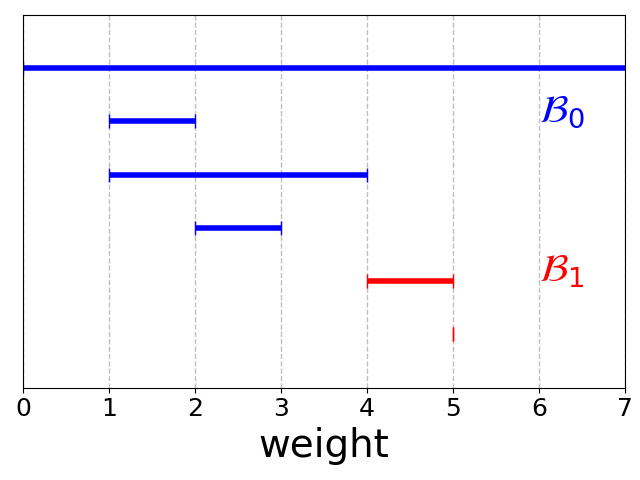}
    \subcaption{}
  \end{minipage}
  \caption{(a) RRM toy model $G$, equivalent to the simplified RRM in Fig.~\ref{fig:simple-graph}. (b) Sequence of subgraphs $G(a)$. (c) Persistence diagram for $X(G)$. (d) Persistence barcode for $X(G)$.}
  \label{fig:rrm-toy-model}
\end{figure}
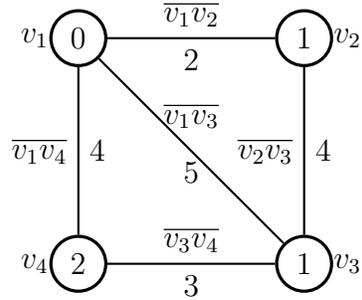
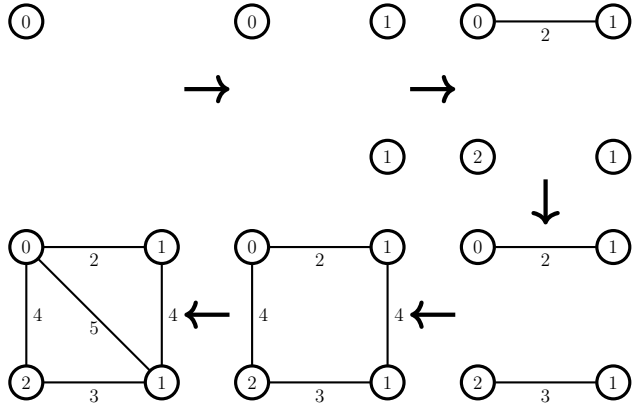
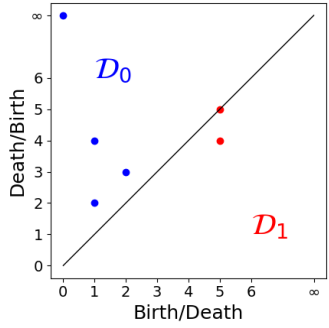
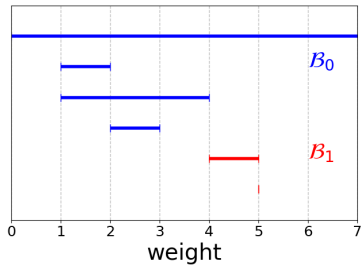

To illustrate the information extracted by the proposed method, its application to the RRM of a toy model was considered (Fig.~\ref{fig:rrm-toy-model}(a)).
The vertices and edges represent the EQ and TS structures, respectively.
For the RRM, $G=(V,E)$, $V=\{{v_1}, {v_2}, {v_3}, {v_4} \}$, and $E=\{ \overline{v_1v_2}, \overline{v_1v_3}, \overline{v_1v_4}, \overline{v_2v_3}, \overline{v_3v_4} \}$.
The weights attributed to the vertices and edges correspond to the relative energies of the corresponding structures with respect to the global minimum energy and are referred to as $w_V(v_1)=0$ and $w_E(\overline{v_1v_2})=2$.
The first generator of $0$-th PH is born at $a=0$, where $V(0)=\{v_1\}$ and $E(0)=\{ \}$. 
Fig.~\ref{fig:rrm-toy-model}(b) depicts the sequence for $G(a)$.
Because this first connected component does not vanish, even when $a \to \infty$, this generator lives forever.
This is plotted as $(b, d)=(0, \infty)$ in $\mathcal{D}_0$ (Fig.~\ref{fig:rrm-toy-model}(c)), whereas it is expressed as the half-line starting from $a=0$ in $\mathcal{B}_0$ (Fig.~\ref{fig:rrm-toy-model}(d)).
Note that the blue (or red) color indicates that the generator is of the 0-th (or 1-st) PH in these representations.
The second and third generators are born simultaneously at $a=1$, where $V(1)=\{v_1, v_2, v_3\}$ and $E(1)=\{ \}$.
At $a=2$, one of these generators dies because edge $\overline{v_1v_2}$ is included in $E(2)$.
This is plotted as $(b, d)=(1, 2)$ in $\mathcal{D}_0$ and is expressed as a line segment between $a=1$ and $2$ in $\mathcal{B}_0$.
Note that the dying generator is not born at $a=0$ because the elder rule is adopted.
Simultaneously, another generator derived from vertex $v_4$ is born.
This dies at $a=3$ by edge $\overline{v_3v_4}$ and is plotted as $(b, d)=(2, 3)$ in $\mathcal{D}_0$.
The other generator born in $a=1$ dies at $a=4$ by the edge $\overline{v_2v_3}$ (or $\overline{v_1v_4}$).
Simultaneously, the generator of the 1-st PH is born because $E(4)=\{ \overline{v_1v_2}, \overline{v_1v_4}, \overline{v_2v_3}, \overline{v_3v_4} \}$ constitutes a ring.
At $a=5$, this ring is split into two rings by edge $\overline{v_1v_3}$ and another generator of the 1-st PH is born.
Note that, if $\varepsilon$ in Eq.~(\ref{eq:clique-weight}) is not zero, the $3$-clique, $\{ v_1, v_2, v_3\}$, e.g., is not included in $X(G)(5)$.
Then, these generators of the 1-st PH die in $a=5+\varepsilon$ where $3$-cliques $\{ v_1, v_2, v_3\}$ and $\{ v_1, v_3, v_4\}$ are included in $X(G)(5+\varepsilon)$.
These are plotted as $(b,d)=(4,5+\varepsilon)$ and $(b,d)=(5,5+\varepsilon)$ in $\mathcal{D}_1$.
The introduction of non-zero $\varepsilon$ enriches the information extracted from PH analysis because when $\varepsilon=0$, only $(b,d)=(4,5)$ is detected.

As the 0-th PH represents the number of connected components corresponding to a certain weight, $a$, an obvious hierarchy exists in the inclusion of vertices.
For example, vertex $v_4$ is unified into vertex $v_3$ for $a=3$.
In the persistence barcode representation, this relation corresponds to a short bar and a longer bar that includes the short bar entirely. 
For example, bar $(b,d)=(2,3)$ is unified into bar $(b,d)=(1,4)$.
Therefore, a long bar corresponds to a \textit{superbasin} \cite{Becker1997_Disconnectivity-graph, Wales2003_EnergyLandscape, Saito1999_WaterDynamics}, defined as a collection of EQs connected by low barriers.
In the persistence diagram representation, this relationship is plotted on the rising left shoulder line.

\begin{figure}
 \begin{minipage}[b]{\linewidth}
  \centering
  \includegraphics[keepaspectratio, scale=1.0]
  {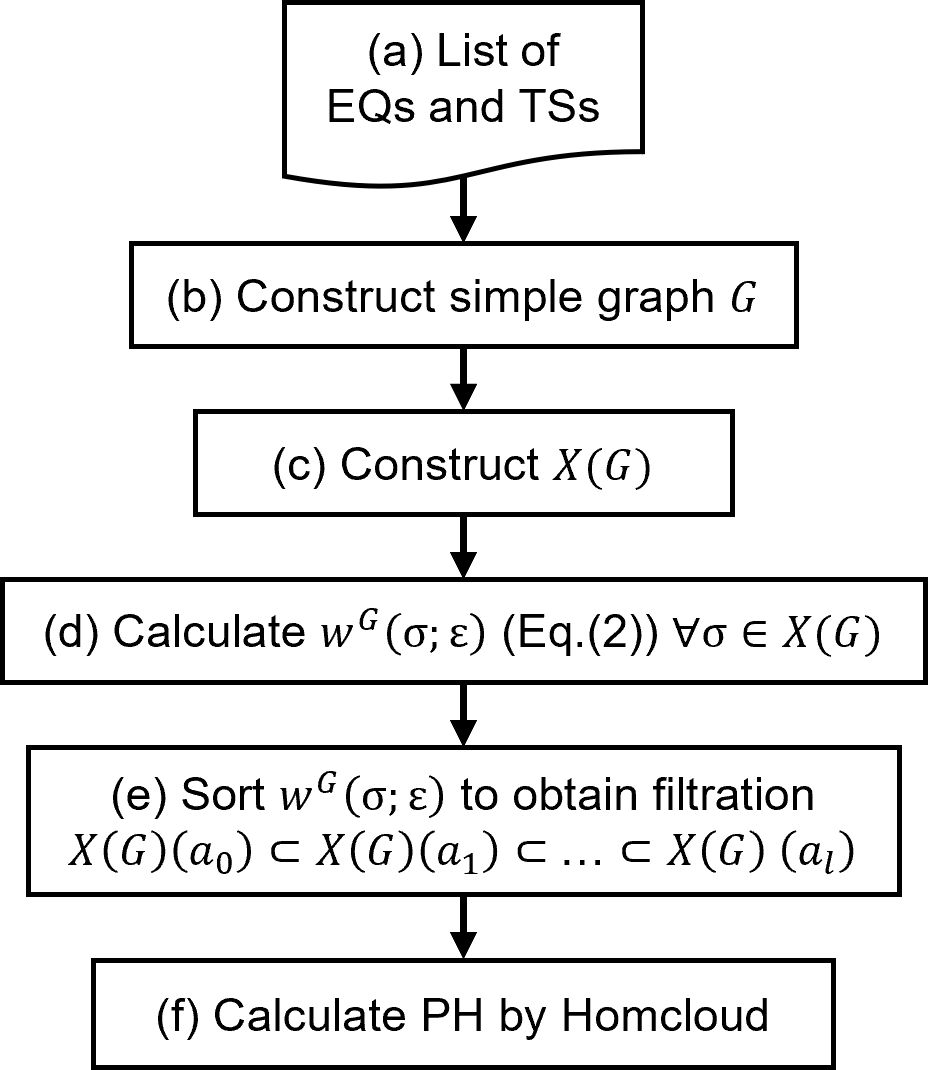}
  \label{flow_chart}
 \end{minipage}
 \caption{Flowchart to obtain the persistence barcode.}
 \label{fig:flow_chart}
\end{figure}

Fig.~\ref{fig:flow_chart} summarizes the flowchart for obtaining the persistence barcode (or equivalently, the persistence diagram) from a list of EQs and TSs.
The code used to obtain the persistence barcode and diagram from the GRRM program output is available on GitHub \cite{github_brian}.

\subsection{Difference from the sublevel set PH of PES}
\label{sec:PH-sublevelset}

\begin{figure}
 \begin{minipage}[b]{0.4\linewidth}
  \centering
  \includegraphics[keepaspectratio, scale=0.15]
  {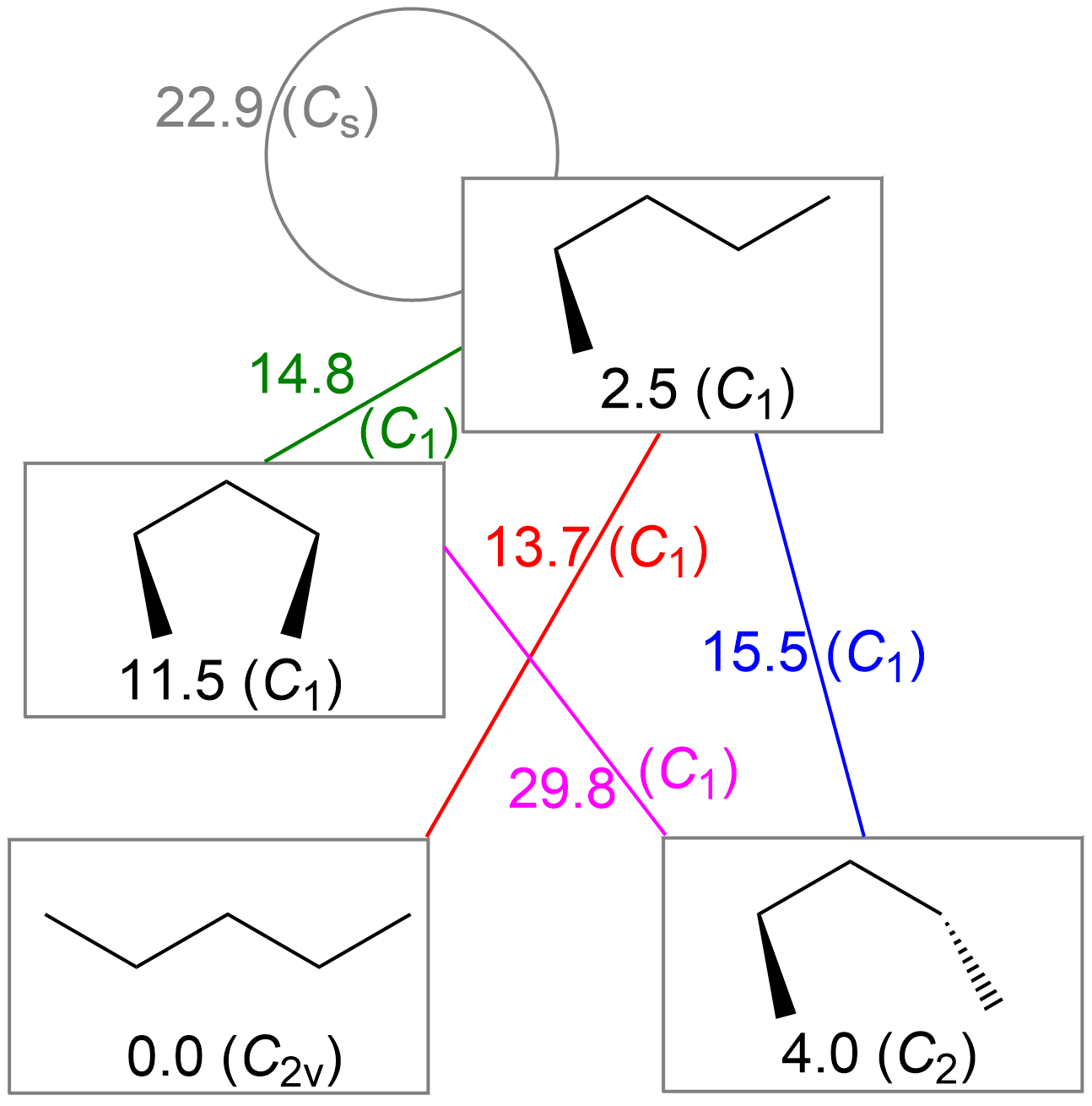}
  \label{nonmirrorRRM}
 \subcaption{}
 \end{minipage}
 \begin{minipage}[b]{0.5\linewidth}
  \centering
  \includegraphics[keepaspectratio, scale=0.13]
  {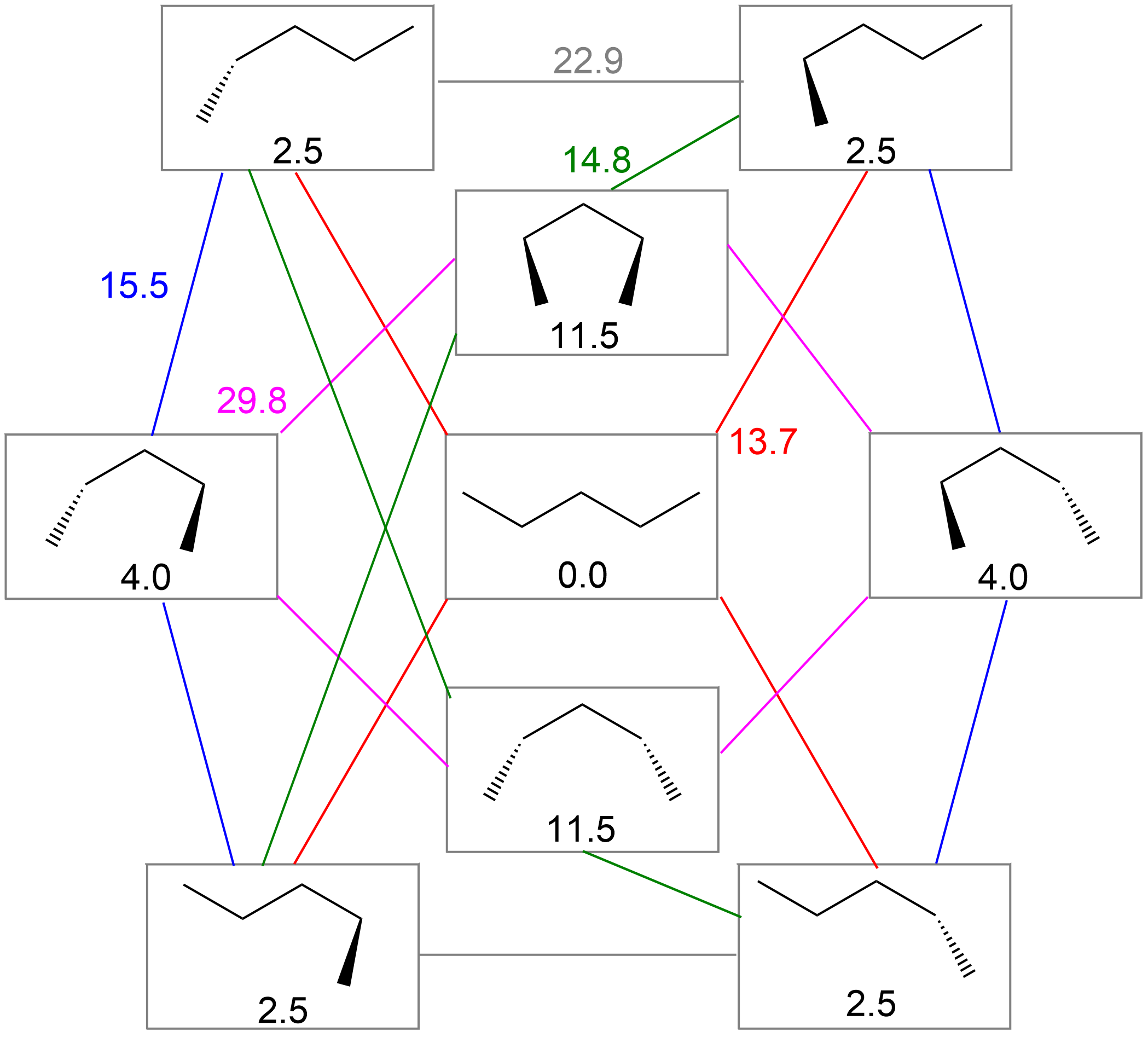}
  \label{mirrorRRM}
 \subcaption{}
 \end{minipage}
 \caption{(a) Partial RRM of \textit{n}-pentane obtained from GRRM-GDSP database and (b) its reconstructed network considering nuclear inversion symmetry.}
 \label{fig:RRM_pentane}
\end{figure}

In Ref.~\citenum{Mirth2021_EL-PH}, Mirth \textit{et al.}~applied sublevel set PH to the two-dimensional model potential of \textit{n}-pentane.
In their sublevel-set PH, a sequence of subspaces of the $(3N-6)$-dimensional PES, in which the energy is smaller than the threshold $a$, is constructed. 
Therefore, all stationary points (not only EQs and TSs but also those with a higher index) in the PES must be located.
Their results revealed that the sublevel-set PH analysis accurately reflects the shape of the PES and produces better descriptors than DG.
In this section, we apply the adjusted weight rank clique filtration to the partial RRM of \textit{n}-pentane, which is constructed based on the GRRM-GDSP database by excluding all bond rearrangement pathways, and compare it with the result of Mirth \textit{et al.}
For comparison, a partial RRM was adopted.
The unmodified partial RRM obtained from GRRM-GDSP \cite{IQCE_GRRM-GDSP} is shown in Fig.~\ref{fig:RRM_pentane}(a).
Although the network size is very compact in this case, it only spans a portion of the full network owing to the nuclear inversion symmetry.
Reconstructing the full network yielded the RRM depicted in Fig.~\ref{fig:RRM_pentane}(b).
The proposed method was applied to this full network.
In this simple case, the full network was reconstructed manually.
A general method for reconstructing a full network by considering the symmetry of the PES is currently under construction and will be reported in a subsequent paper. \cite{Teramoto2023_CNPI}

\begin{figure}[H]
  \begin{minipage}[b]{0.3\linewidth}
  \centering
  \includegraphics[keepaspectratio,width=\textwidth]
  {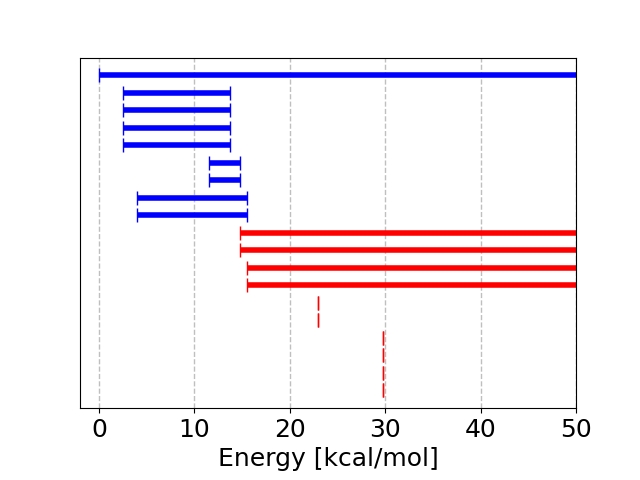}
 \subcaption{}
 \end{minipage}
 \begin{minipage}[b]{0.3\linewidth}
  \centering
  \includegraphics[keepaspectratio, width=\textwidth]
  {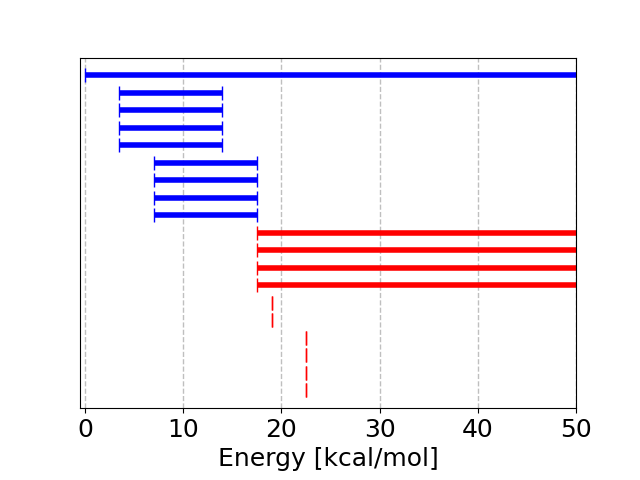}
 \subcaption{}
 \end{minipage}
  \begin{minipage}[b]{0.3\linewidth}
  \centering
  \includegraphics[keepaspectratio, width=\textwidth]
  {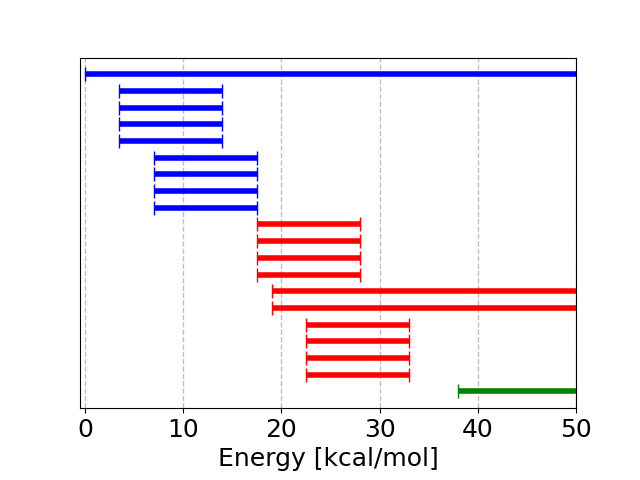}
 \subcaption{}
 \end{minipage} 
 \caption{Persistence barcodes of \textit{n}-pentane obtained using (a,b) adjusted weight rank clique filtration for (a) DFT level energy and (b) model potential
 and (c) sublevel set PH for model potential (from Ref.~\citenum{Mirth2021_EL-PH}, green bar corresponds to the 2-nd persistence barcode) }
 \label{fig:PB_pentane}
\end{figure}

Figure \ref{fig:PB_pentane}(a) depicts the persistence barcode obtained from the adjusted weight rank clique filtration of the RRM (Fig.~\ref{fig:RRM_pentane}(b)).
Figure \ref{fig:PB_pentane}(b) is the same figure, but for the model potential adopted in Ref.~\citenum{Mirth2021_EL-PH}.
For comparison, the persistence barcode obtained by filtration based on the sublevel sets by Mirth \textit{et al.} was reprinted (Fig.~\ref{fig:PB_pentane}(c)).
Here, we denote the vertex and edge with weights $w$ as $v(w)$ and $e(w)$, respectively (when expressed as $V(w)$ and $E(w)$ with a capital letter, they represent the set collected up to the threshold $w$).
For the 0-th persistence barcode ($\mathcal{B}_0$), the bars in Figs.~\ref{fig:PB_pentane}(a) and (b) show qualitative agreement, although the bar lengths were different.
However, the bars are identical for Figs.~\ref{fig:PB_pentane}(b) and (c).
This clearly shows that the current filtration can extract the same information as the sublevel-set filtration if the adopted potential is the same.
Specifically, for the first four bars other than the top semi-infinite bar, the 0-th persistence barcode corresponds to $v(2.5)$ and $e(13.7)$ (red edges) in Fig.~\ref{fig:RRM_pentane}(b).
In the case of the model potential conducted by Mirth \textit{et al.}, the energies of $v(11.5)$ and $e(14.8)$ (green edges) coincide with those of $v(4.0)$ and $e(15.5)$ (blue edges), respectively, owing to the lack of steric repulsion.
Therefore, the next four bars in $\mathcal{B}_0$ correspond to the green edges and their associated vertices.
By accounting for steric repulsion, which was not considered in the model potential, the degeneracy of the bars was partially resolved using the proposed method.
As for the 1-st persistence barcode ($\mathcal{B}_1$), the number of bars is identical among Figs.~\ref{fig:PB_pentane}(a), (b), and (c), while their births in Fig.~\ref{fig:PB_pentane}(a) differ from those shown in Figs.~\ref{fig:PB_pentane}(b) and (c) owing to the difference in the adopted potential.
For $\mathcal{B}_1$ in Fig.~\ref{fig:PB_pentane}(a), the first two bars and the next two bars correspond to cycles $(v(0.0), v(2.5), v(11.5), v(2.5))$ and $(v(0.0), v(2.5), v(4.0), v(2.5))$, respectively, which also degenerate in Fig.~\ref{fig:PB_pentane}(b).
In Fig.~\ref{fig:PB_pentane}(c), these bars die because of the second-order saddle point, although they do not die in Figs.~\ref{fig:PB_pentane}(a) and (b) owing to the lack of a direct pathway between $v(0.0)$ and $v(11.5)/v(4.0)$ or $v(2.5)$ and another $v(2.5)$.
The next two bars correspond to cycles $(v(0.0), v(2.5), v(2.5))$.
By Figs.~\ref{fig:PB_pentane}(a) and (b), the length of these bars is $\varepsilon$ because they are 3-cliques.
In contrast, in Fig.~\ref{fig:PB_pentane}(c), the bars do not die because the configuration space of the model potential adopted by Mirth \textit{et al.} is a two-dimensional torus \cite{Mirth2021_EL-PH}.
For the final four bars corresponding to cycles $(v(2.5), v(4.0), v(11.5))$, those depicted in Fig.~\ref{fig:PB_pentane}(c) die by the second-order saddle point, whereas the lengths in Figs.~\ref{fig:PB_pentane}(a) and (b) are $\varepsilon$ because they are 3-cliques.
Note that in the low-temperature limit, or in the range where the transition state theory is valid, because the second-order saddle point does not affect the kinetics, the difference between Figs.~\ref{fig:PB_pentane}(b) and (c) does not affect the interpretation of the chemical kinetics. 
The death of the bar in the 1-st persistence barcode is just a descriptor defined as a quantity characterizing the PES or RRM.

\subsection{Computational Details}

RRMs of the metal nanoclusters were constructed using the SC-AFIR method implemented in the GRRM program.
The force parameter $\gamma=300$ kJ mol$^{-1}$ is adopted during the search for the reaction path.
Electronic structure calculations were performed using the Gaussian16 program \cite{g16} with the Perdew--Burke--Ernzerhof (PBE) exchange-correlation functional \cite{Perdew1996_PBE} and the LanL2DZ basis set associated with the effective core potential \cite{Hay1985_LanL2DZ}.
The RRMs of the organic molecules were obtained from Prof.~K.~Ohno at the Institute for Quantum Chemical Exploration (IQCE) and are available on the GRRM-GDSP website \cite{IQCE_GRRM-GDSP}.
The level of the electronic structure calculations was designated as B3LYP/6-31G(d).
The RRM for the Claisen rearrangement reported in Ref.~\citenum{Nagahata2016_Claisen}, where the geometry optimizations were performed using the M06-2X exchange-correlation functional \cite{Zhao2008_M06-family} with the 6-311+G(2d,p) basis set \cite{Krishnan1980_6-311G2dp, Clark1983_3-21+G_diffuse} and the single-point energies were recalculated using the CCSD(T) method alongside the jun-cc-pVTZ basis set \cite{dunning1989gaussian, papajak2011convergent}, was provided by Prof.~Maeda and Dr.~Nagahata.
A program to reproduce our methods is available in Ref.~\citenum{github_brian}.

\section{Results and Discussion}
\label{sec:Results} 

In this section, we describe the chemical interpretation of the information obtained from the persistence barcode (or equivalently, the persistence diagram). 

\subsection{Persistence barcode/diagram for RRMs of metal nanoclusters}

\begin{figure*}
  \begin{minipage}[b]{0.95\linewidth}
    \centering
    \includegraphics[keepaspectratio, scale=0.35]{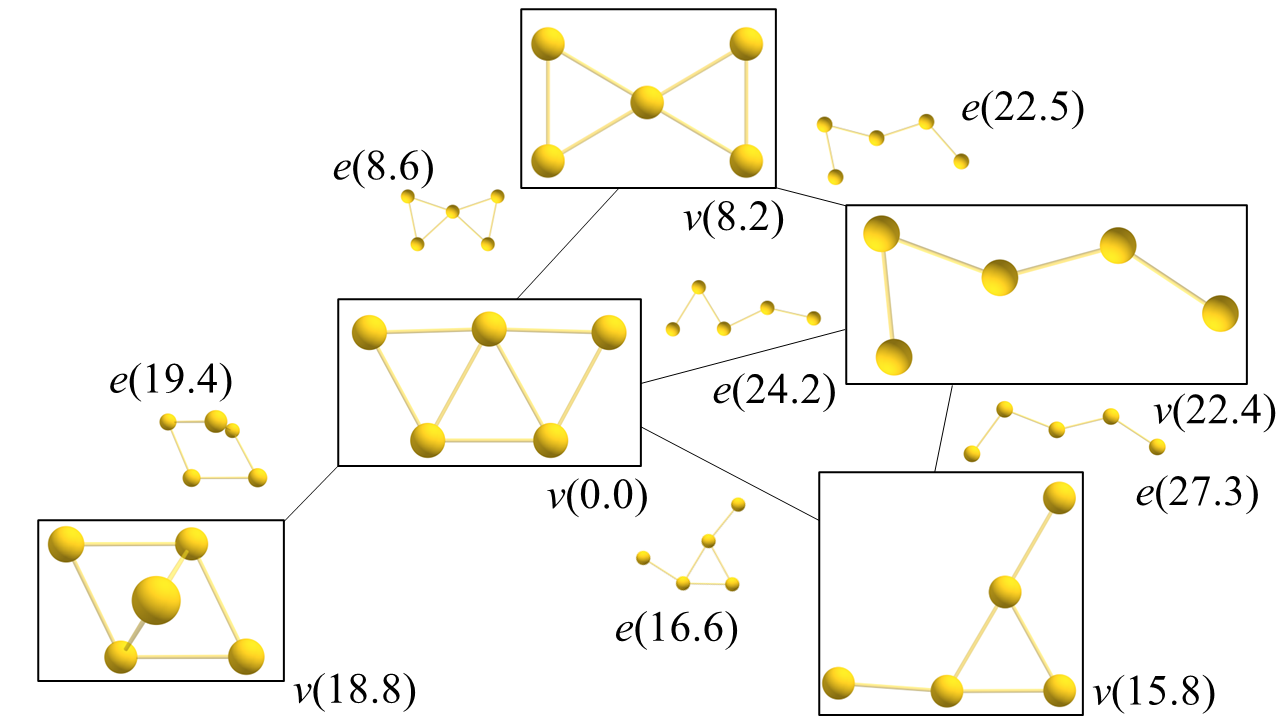}
    \subcaption{}
  \end{minipage}

  \begin{minipage}[b]{0.45\linewidth}
    \centering
    \includegraphics[keepaspectratio, scale=0.45]{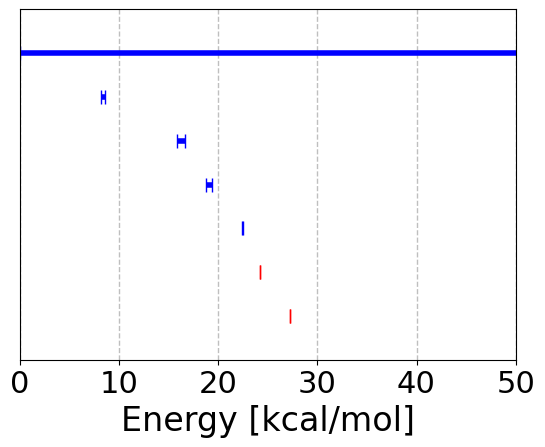}
    \subcaption{}
  \end{minipage}
  \begin{minipage}[b]{0.45\linewidth}
    \centering
    \includegraphics[keepaspectratio, scale=0.45]{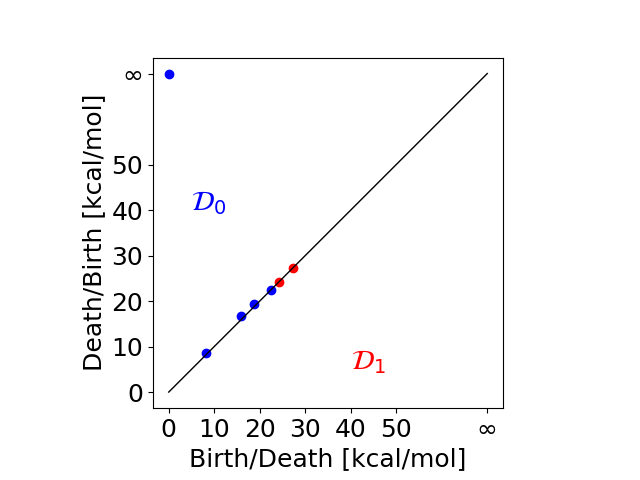}
    \subcaption{}
  \end{minipage}
  \caption{(a) RRM of Au$_5$ with corresponding structures, (b) its persistence barcode, and (c) its persistence diagram.}
  \label{fig:Au5}
  \end{figure*}

First, we focus on the RRM of the Au$_5$ cluster, which was investigated in detail in Ref.~\citenum{Harabuchi2015_Au5-bifurcation}, and is illustrated in Fig.~\ref{fig:Au5} (a).
The persistence barcode constructed using the proposed method is shown in Fig.~\ref{fig:Au5} (b).
The number attributed to each vertex or edge is the relative energy in kcal mol$^{-1}$.
In this section, each vertex, $v$, or edge, $e$, is designated by its energy value; for instance, $v(0.0)$ refers to the vertex with an energy value of 0.0 kcal mol$^{-1}$ corresponding to the W-shaped $C_{2v}$ structure.
In this RRM, $v(0.0)$ is the global minimum and is connected to all other vertices.
At $a=8.2$ (henceforth, the unit of energy is omitted), $v(8.2)$ corresponding to the X-shaped $D_{2h}$ structure appears, but is disconnected from the former vertex, $v(0.0)$.
They are merged at $a=8.6$, leading to a generator of the 0-th PH with $(b,d)=(8.2, 8.6)$.
This generator of the 0-th PH indicates that a potential basin with a minimum energy of 8.2 kcal mol$^{-1}$ is incorporated within another potential basin (in this case, $v(0.0)$) if the energy parameter $a$ is raised to 8.6 kcal mol$^{-1}$.
Similarly, the generators of the 0-th PH with $(b, d)=(15.8, 16.6)$, $(18.8, 19.4)$, and $(22.4, 22.5)$ were obtained from this RRM.
The lengths of the bars of these generators (i.e., $d-b$) were significantly small.
This implies that the basins corresponding to these EQs [$v(8.2)$, $v(15.8)$, $v(18.8)$, and $v(22.4)$] are very shallow.
It should be noted that this persistence barcode (and persistence diagram) cannot directly provide information for identifying the basin to which the merged basin is incorporated---this can be expressed in the disconnectivity graph. 
At $a=24.2$, $v(0.0)$ is directly connected to $v(22.4)$, whereas they are already connected via $v(8.2)$ at $a=22.5$.
The emergence of two different routes (i.e., stepwise and concerted pathways) causes the birth of a generator of the 1-st PH.
Because these three vertices share a complete connection at $a=24.2$, the generator dies immediately at $a=24.2 + \varepsilon$.
The case of the generator with $(b,d)=(27.3, 27.3+\varepsilon)$, which originates from vertices ${v(0.0), v(15.8), v(22.4)}$, is similar.
For an intuitive understanding, we provide a one-to-one correspondence between the element of the RRM and the bar, as depicted in Fig.~S1 in the Supporting Information.

\begin{figure}[H]
 \begin{minipage}[b]{0.49\linewidth}
 \centering
  \includegraphics[keepaspectratio, scale=0.26]
  {images/PD_and_PB/PBM/Au5_barcode.png}
 \subcaption{Au$_5$}
 \end{minipage}
 \begin{minipage}[b]{0.49\linewidth}

  \centering
  \includegraphics[keepaspectratio, scale=0.26]
  {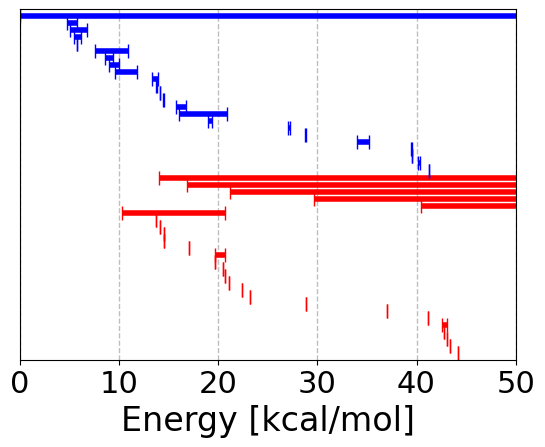}
 \subcaption{Au$_7$}
 \end{minipage}
 \begin{minipage}[b]{0.49\linewidth}
 
  \centering
  \includegraphics[keepaspectratio, scale=0.26]
  {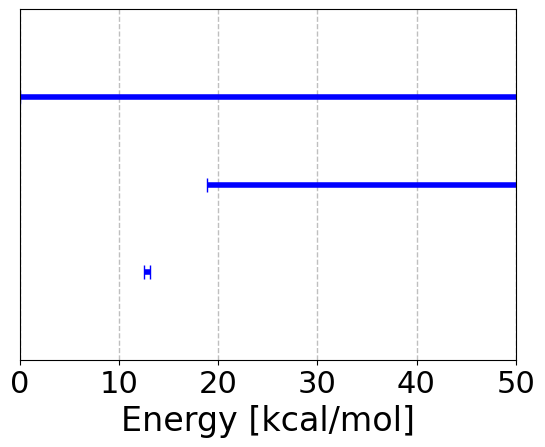}
 \subcaption{Ag$_5$}
 \end{minipage}
 \begin{minipage}[b]{0.49\linewidth}
  \centering
  \includegraphics[keepaspectratio, scale=0.26]
  {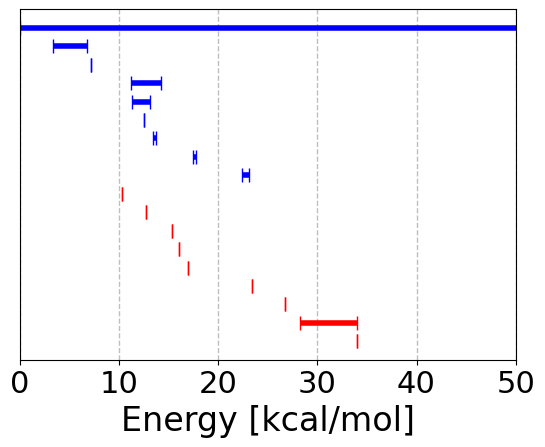}
 \subcaption{Ag$_7$}
 \end{minipage}
 \begin{minipage}[b]{0.49\linewidth}
  \centering
  \includegraphics[keepaspectratio, scale=0.26]
  {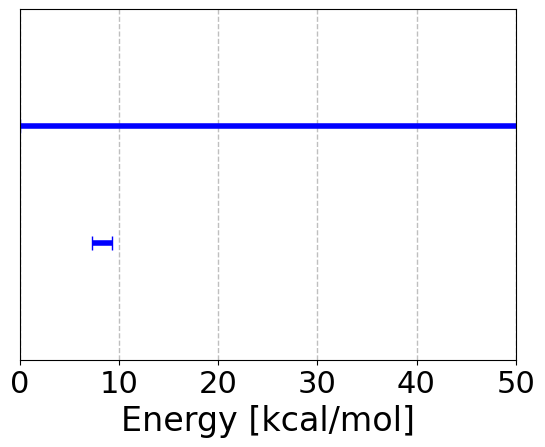}
 \subcaption{Cu$_5$}
 \end{minipage}
 \begin{minipage}[b]{0.49\linewidth}
  \centering
  \includegraphics[keepaspectratio, scale=0.26]
  {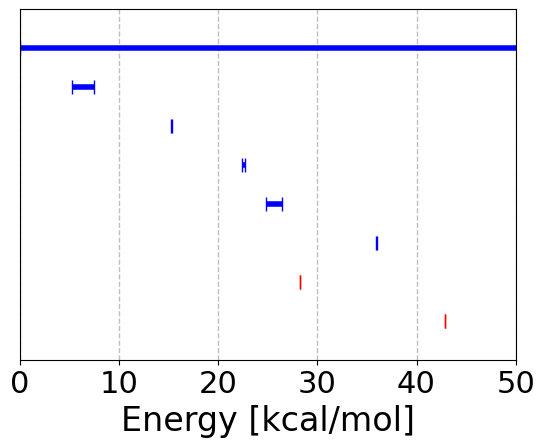}
 \subcaption{Cu$_7$}
 \end{minipage}
 \caption{Persistence barcodes for RRMs of (a) Au$_5$, (b) Au$_7$, (c) Ag$_5$, (d) Ag$_7$, (e) Cu$_5$, and (f) Cu$_7$ systems.}
 \label{fig:coin-single-PH}
\end{figure}

Figure \ref{fig:coin-single-PH} depicts the persistence barcodes for the RRMs of single-element coinage metal clusters with five and seven atoms.
If the number of atoms is fixed, the number of generators of PH increases towards the bottom of the periodic table
(the number of bars and the average and standard deviation (SD) of bar lengths for each barcode are summarized in Table S1 in the Supporting Information).
As the number of generators of the 0-th PH is related to the number of EQ geometries, this tendency can be interpreted to be indicative of the hardness of the element, i.e., softer elements have more generators. 
For the same element, the number of generators increases as the number of atoms increases.
Also, the number of long bars in the persistence barcode tends to increase with an increase in the number of atoms.
This implies that the number of metastable cluster structures that can persist over a lengthy period increases as the cluster size increases.

\begin{figure}
 \begin{minipage}[b]{0.49\linewidth}
  \centering
  \includegraphics[keepaspectratio, scale=0.26]
  {images/PD_and_PB/PBM/Au5_barcode.png}
 \subcaption{Au$_5$}\label{Au5}
 \end{minipage}
 \begin{minipage}[b]{0.49\linewidth}
  \centering
  \includegraphics[keepaspectratio, scale=0.26]
  {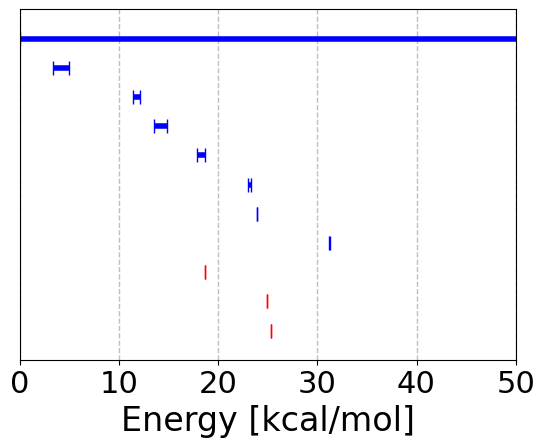}
 \subcaption{Au$_4$Cu}\label{Au4Cu}
 \end{minipage}
 \begin{minipage}[b]{0.49\linewidth}
  \centering
  \includegraphics[keepaspectratio, scale=0.26]
  {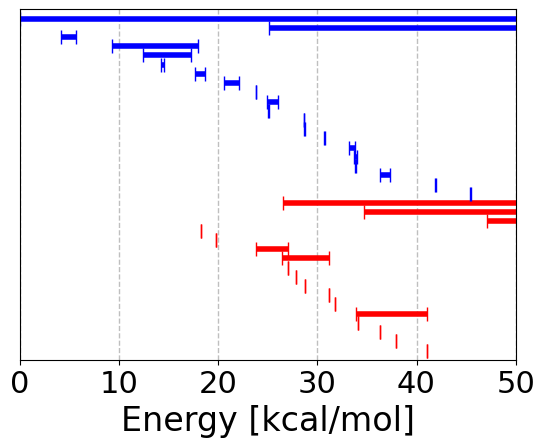}
 \subcaption{Au$_3$Cu$_2$}\label{Au3Cu2}
 \end{minipage}
 \begin{minipage}[b]{0.49\linewidth}
  \centering
  \includegraphics[keepaspectratio, scale=0.26]
  {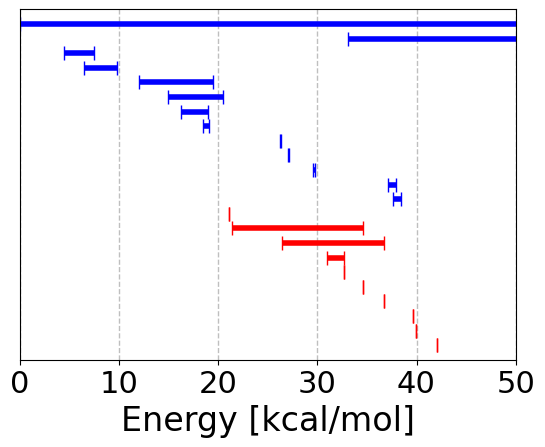}
 \subcaption{Au$_2$Cu$_3$}\label{Au2Cu3}
 \end{minipage}
 \begin{minipage}[b]{0.49\linewidth}
  \centering
  \includegraphics[keepaspectratio, scale=0.26]
  {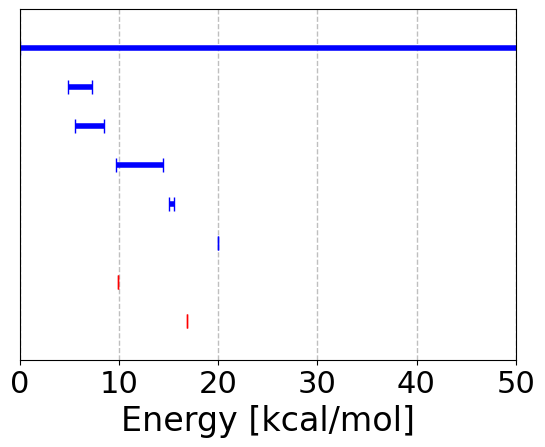}
 \subcaption{AuCu$_4$}\label{AuCu4}
 \end{minipage}
 \begin{minipage}[b]{0.49\linewidth}
  \centering
  \includegraphics[keepaspectratio, scale=0.26]
  {images/PD_and_PB/PBM/Cu5_barcode.png}
 \subcaption{Cu$_5$}\label{Cu5}
 \end{minipage}
 \caption{Persistence barcodes for RRMs of (a) Au$_5$, (b) Au$_4$Cu, (c) Au$_3$Cu$_2$, (d) Au$_2$Cu$_3$, (e) AuCu$_4$, and (f) Cu$_5$.} \label{fig:coin-double-PH}
\end{figure}

Figure \ref{fig:coin-double-PH} depicts the persistence barcodes for the RRMs of the five-atomic alloy clusters of Au and Cu.
The results corresponding to all possible mixing patterns (i.e., Au$_5$, Au$_4$Cu, Au$_3$Cu$_2$, Au$_2$Cu$_3$, AuCu$_4$, and Cu$_5$) are presented.
As the mixing ratio becomes more uniform (i.e., $X_5 \to X_4Y \to X_3Y_2$), the number of generators of the PH increases.
This is related to the number of permutation isomers, which increases as the mixing ratio becomes more uniform. 
Interestingly, the average length of the bars in the barcode increases as the mixing ratio becomes more uniform, indicating an increase in the number of metastable cluster structures persisting over long intervals.

\begin{figure}[H]
 \begin{minipage}[b]{0.49\linewidth}
  \centering
  \includegraphics[keepaspectratio, scale=0.26]
  {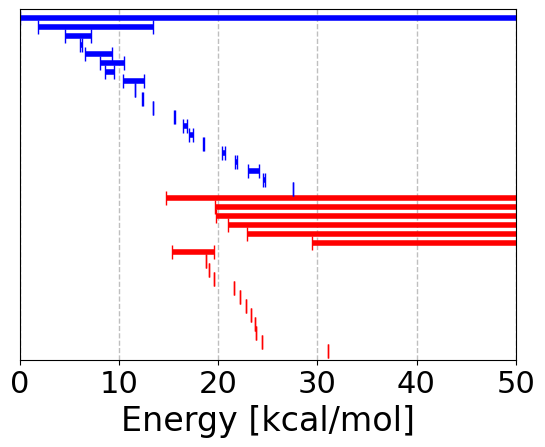}
 \subcaption{Au$_3$Ag$_2$}
 \end{minipage}
 \begin{minipage}[b]{0.49\linewidth}
  \centering
  \includegraphics[keepaspectratio, scale=0.26]
  {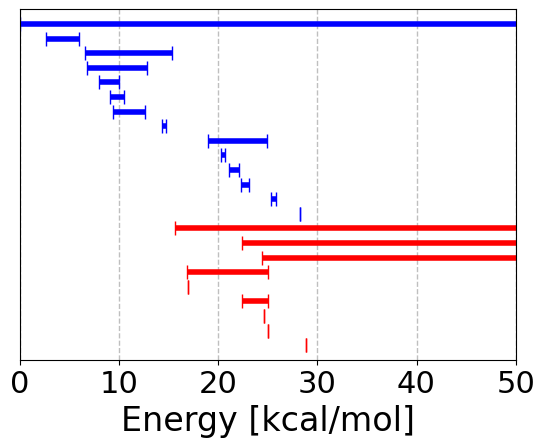}
 \subcaption{Au$_2$Ag$_3$}
 \end{minipage}
 \begin{minipage}[b]{0.49\linewidth}
  \centering
  \includegraphics[keepaspectratio, scale=0.26]
  {images/PD_and_PB/PBM/Au3Cu2_barcode.png}
 \subcaption{Au$_3$Cu$_2$}
 \end{minipage}
 \begin{minipage}[b]{0.49\linewidth}
  \centering
  \includegraphics[keepaspectratio, scale=0.26]
  {images/PD_and_PB/PBM/Au2Cu3_barcode.png}
 \subcaption{Au$_2$Cu$_3$}
 \end{minipage}
 \begin{minipage}[b]{0.49\linewidth}
  \centering
  \includegraphics[keepaspectratio, scale=0.26]
  {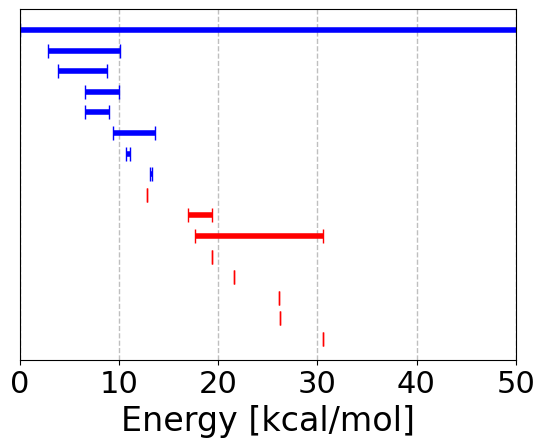}
 \subcaption{Ag$_3$Cu$_2$}
 \end{minipage}
 \begin{minipage}[b]{0.49\linewidth}
  \centering
  \includegraphics[keepaspectratio, scale=0.26]
  {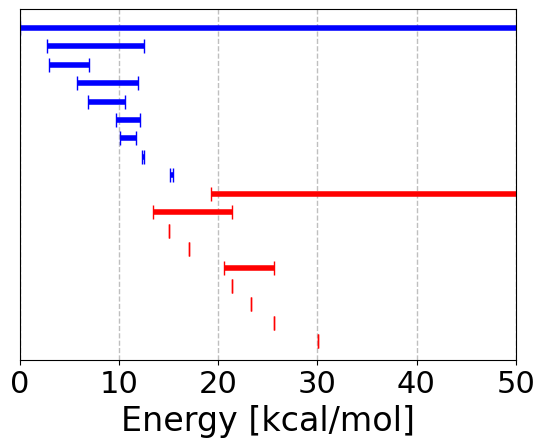}
 \subcaption{Ag$_2$Cu$_3$}
 \end{minipage}
 \caption{Persistence barcodes for RRMs of (a) Au$_3$Ag$_2$, (b) Au$_3$Cu$_2$, (c) Ag$_3$Cu$_2$, (d) Au$_2$Ag$_3$, (e) Au$_2$Cu$_3$, and (f) Ag$_2$Cu$_3$}\label{fig:coin-X3Y2-PH}
\end{figure}

Figure \ref{fig:coin-X3Y2-PH} compares the persistence barcodes for the RRMs of $X_3Y_2$ ($X, Y$ = Cu, Ag, or Au) corresponding to different elements. 
For the same combination of elements (i.e., $X_3Y_2$ and $X_2Y_3$), the persistence barcodes tend to exhibit similar trends. 
For the combination of Au and Cu, the generators of PH appear over a wide energy range, extending up to 50 kcal mol$^{-1}$, while for other combinations, they mostly appear at up to 30 kcal mol$^{-1}$.
The number of generators is comparatively lower for the combination of Ag and Cu than for the others.
On the contrary, for the combination of Au and Ag, the number of generators, especially those of the 1-st PH living forever, is higher than those of the others.
This suggests that the analysis can be used to extract the affinity of two metal elements or the properties of metal alloys. 

\begin{figure*}
  \begin{minipage}[b]{0.95\linewidth}
    \centering
    \includegraphics[keepaspectratio, scale=0.45]{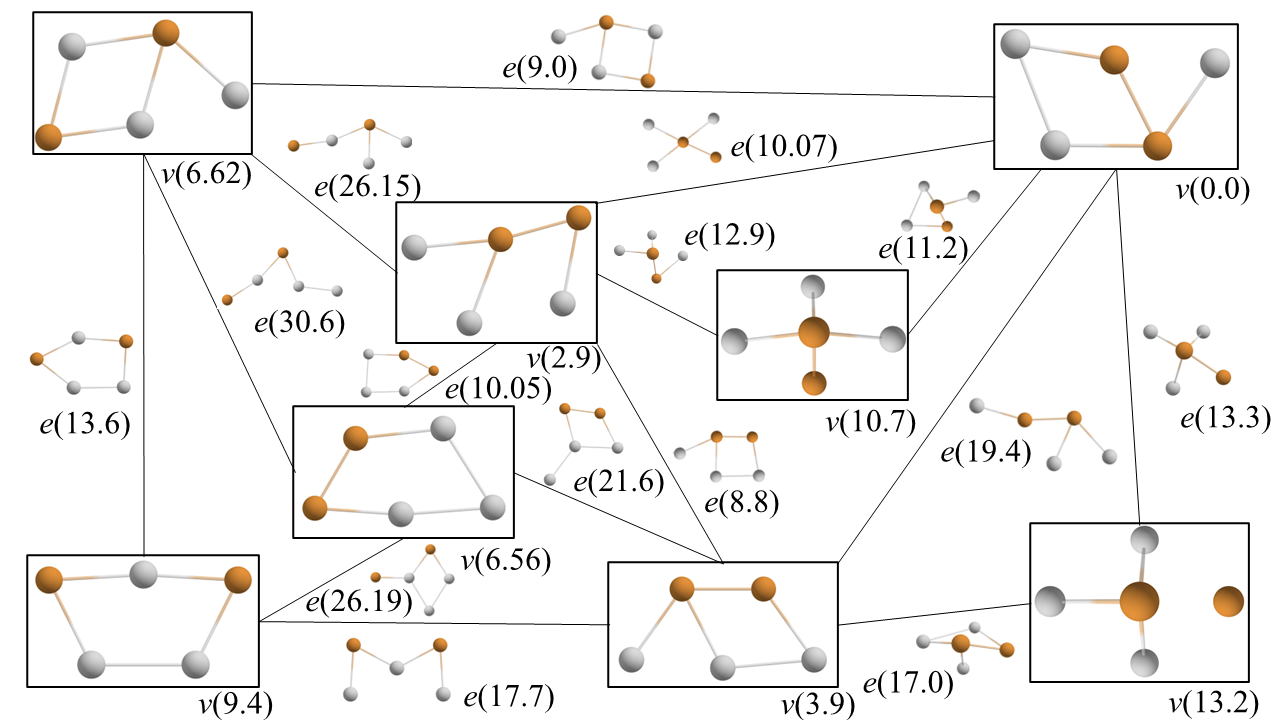}
  \end{minipage}
  \caption{RRM of Ag$_3$Cu$_2$ with their structures. This corresponds to Fig.~\ref{fig:coin-X3Y2-PH} (e)} 
  \label{fig:Ag3Cu2-RRM}
\end{figure*}

The RRM for Ag$_3$Cu$_2$ is depicted in Fig.~\ref{fig:Ag3Cu2-RRM}.
Vertices $v(2.9)$ and $v(3.9)$ appear at $a=2.9$ and $a=3.9$, respectively.
Each of $v(0.0)$, $v(2.9)$, and $v(3.9)$ belongs to a different connected component from the others up to $a=8.8$.
Correspondingly, the generators of the 0-th PH are born at $a=2.9$ and $a=3.9$.
At $a=8.8$, $e(8.8)$ connects $v(2.9)$ and $v(3.9)$ and the generator born at $a=3.9$ dies based on the elder rule.
In this case, unlike in the case of Au$_5$, the persistence barcode does not uniquely determine whether $v(3.9)$ is connected to $v(0.0)$ or $v(2.9)$.
At $a=12.9$, a loop consisting of three vertices ($v(0.0)$, $v(2.9)$, and $v(10.7)$) appears, and a generator of the 1-st PH is born.
Because this loop is a $3$-clique, this generator dies at $a=12.9 + \varepsilon$.
At $a=17.0$, a loop consisting of four vertices ($v(0.0)$, $v(2.9)$, $v(3.9)$, and $v(13.2)$) appears, and a generator of the 1-st PH is born.
The generator does not die immediately.
At $a=19.4$, $v(0.0)$ is directly connected to $v(3.9)$ via $e(19.4)$ and ${v(0.0), v(2.9), v(3.9)}$ and ${v(0.0), v(3.9), v(13.2)}$ constitute two $3$-cliques.
Then, another generator of the 1-st PH is born at $a=19.4$ and the generators of the 1-st PH die at $a=19.4 + \varepsilon$.
As discussed above, the generator of the 1-st PH provides information regarding the detour in the reaction pathway.

\begin{figure}[H]
  \begin{minipage}[b]{0.95\linewidth}
    \centering
    \includegraphics[keepaspectratio, scale=0.45]{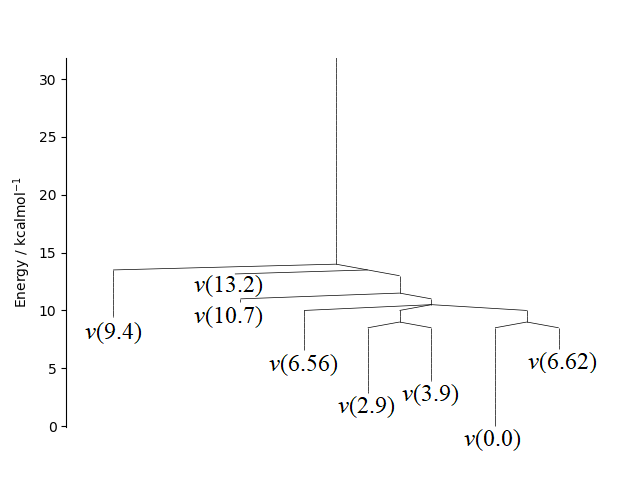}
  \end{minipage}
  \caption{DG for the RRM of Ag$_3$Cu$_2$. DG provides inclusion relationships between the two potential basins. However, DG cannot provide information on loop structures.} 
  \label{fig:Ag3Cu2-DG}
\end{figure}

For comparison, the DG of the RRM of Ag$_3$Cu$_2$ (Fig.~\ref{fig:Ag3Cu2-RRM}), which was illustrated using PyConnect \cite{Smeeton2014_PyConnect}, is depicted in Fig.~\ref{fig:Ag3Cu2-DG}.
DG complements the persistence barcode of the 0-th PH by providing inclusion relationships between the two potential basins.
For example, it is evident from the DG that vertex $v(3.9)$ is connected to $v(2.9)$ at $a=8.8$.
However, as mentioned previously, DG cannot provide higher-order information about connections, such as loop structures.
Both DG and persistent homology methods provide descriptors regarding the topology of the RRM.

\subsection{Persistence diagram for RRMs of organic molecules}

In this subsection, the proposed method is applied to RRMs of organic molecules.
The diagram representation of PH is used because the number of generators in this case can be significantly larger than that of the metal clusters. 
Again, note that the bar length in $\mathcal{B}_0$ (or $\mathcal{B}_1$) corresponds to the vertical (or horizontal) distance from the diagonal line in $\mathcal{D}_0$ (or $\mathcal{D}_1$).

\begin{figure}[H]
 \begin{minipage}[b]{0.4\linewidth}
  \centering
  \includegraphics[keepaspectratio, width=\textwidth]
  {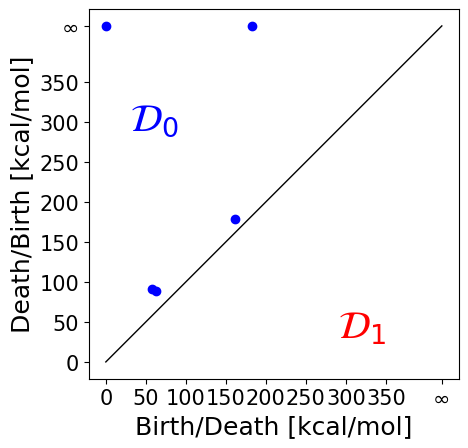}
 \subcaption{H$_2$CO}\label{H2CO}
 \end{minipage}
 \begin{minipage}[b]{0.4\linewidth}
  \centering
  \includegraphics[keepaspectratio, width=\textwidth]
  {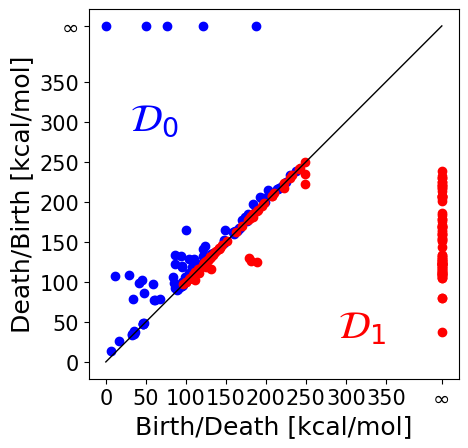}
 \subcaption{H$_4$C$_2$O$_2$}\label{H4C2O2}
 \end{minipage}\\
 \caption{Persistence diagrams for RRMs of (a) H$_2$CO and (b) H$_4$C$_2$O$_2$}
 \label{fig:PD_H2CO}
\end{figure}

First, the size dependencies of the persistence diagrams are compared corresponding to identical chemical element ratios.
Fig.~\ref{fig:PD_H2CO} depicts persistence diagrams for the RRMs of H$_2$CO and H$_4$C$_2$O$_2$.
The number of generators of both the 0-th and 1-st PH significantly increases as the number of atoms is doubled.
However, the plots appearing in the persistence diagram of H$_2$CO can be found at positions like those of H$_4$C$_2$O$_2$.
In addition, the number of plots, apart from the diagonal line, increases as the system size increases.
This trend was also confirmed for the metal nanocluster (Fig.~\ref{fig:coin-single-PH}).

\begin{figure}[H]
 \begin{minipage}[b]{0.3\linewidth}
  \centering
  \includegraphics[keepaspectratio, width=\textwidth]
  {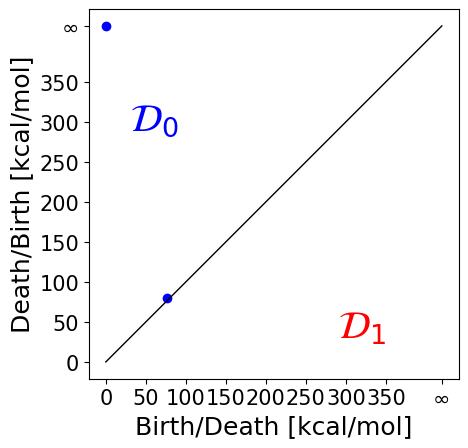}
 \subcaption{H$_4$C$_2$}\label{H$_4$C$_2$}
 \end{minipage}
 \begin{minipage}[b]{0.3\linewidth}
  \centering
  \includegraphics[keepaspectratio, width=\textwidth]
  {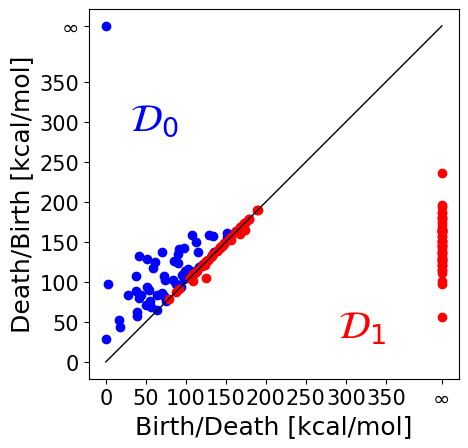}
 \subcaption{H$_2$C$_2$N$_2$}\label{H2C2N2}
 \end{minipage}
 \begin{minipage}[b]{0.3\linewidth}
  \centering
  \includegraphics[keepaspectratio, width=\textwidth]
  {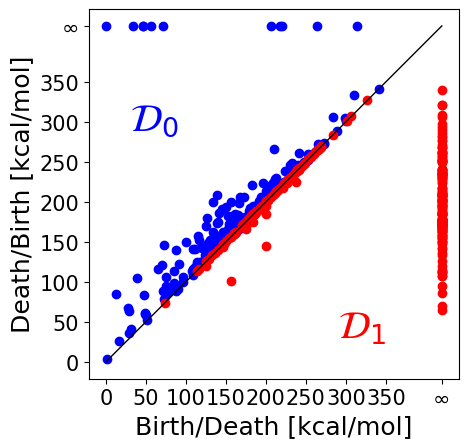}
 \subcaption{HC$_2$NO$_2$}\label{HC2NO2}
 \end{minipage}
 \caption{Persistence diagrams for RRMs of (a) H$_4$C$_2$, (b) H$_2$C$_2$N$_2$, and (c) HC$_2$NO$_2$}
 \label{fig:PD_organic6}
\end{figure}

Next, the persistence diagrams for RRMs of H$_4$C$_2$, H$_2$C$_2$N$_2$, and HC$_2$NO$_2$ are compared, where identical numbers of atoms but different numbers of chemical elements are considered (Fig.~\ref{fig:PD_organic6}).
For H$_4$C$_2$, there are only 2 EQs, namely, ethylene (H$_2$C$=$CH$_2$) and HCCH$_3$---the energy of the latter is 76.1 kcal mol$^{-1}$ higher than that of ethylene.
Therefore, in this case, there exists only one generator of the 0-th PH (except for the obvious one with $(b,d)=(0,\infty)$) corresponding to the TS between these two EQs and there is no generator of the 1-st PH.
As the number of elements increases, the range of the plots gradually expands.
Simultaneously, the numbers of generators of both the 0-th and 1-st PH increase significantly.
In particular, the number of plots far from the diagonal increases, which can be confirmed from the average bar length in $\mathcal{B}_1$) provided in Table S1 in the Supporting Information, while the trend is not obvious in $\mathcal{D}_0$ for H$_2$C$_2$N$_2$ and HC$_2$NO$_2$.
This indicates that the number of stable isomers increases as the number of elements increases, even when the number of atoms is maintained constant.
However, it should be noted that nuclear permutation and inversion symmetry is not considered in the current application.
As discussed in Sec.~\ref{sec:PH-sublevelset}, the persistence diagram changes when symmetrically degenerated vertices and edges are resolved. 
In addition, in $\mathcal{D}_1$, the plot with $d = \infty$ increases significantly as the number of elements increases.
This may indicate that the two different reaction pathways connecting a pair of EQs do not primarily consist of direct (concerted) and two-step pathways in RRMs of organic molecular systems compared to those of metal clusters. 

\begin{figure*}
  \begin{minipage}[b]{0.9\linewidth}
    \centering
    \includegraphics[keepaspectratio, scale=0.40]{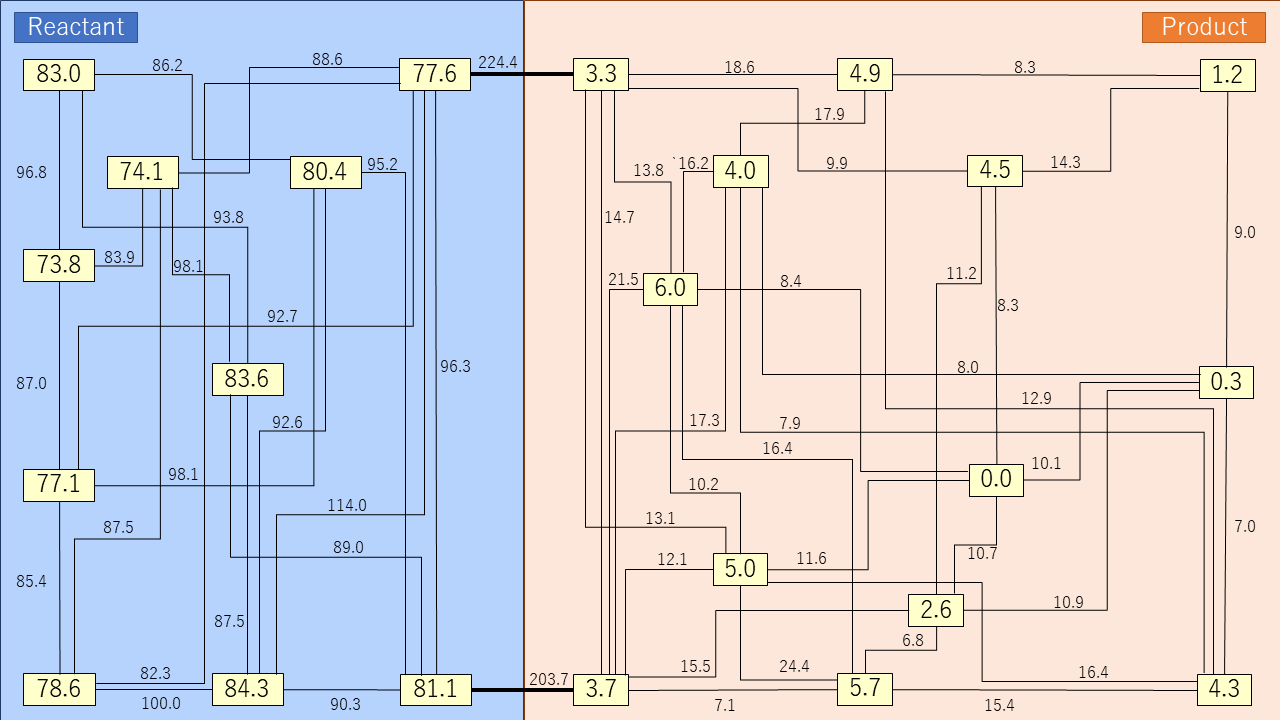}
  \subcaption{}
  \end{minipage}\\
  \begin{minipage}[b]{0.50\linewidth}
    \centering
    \includegraphics[keepaspectratio,scale=0.5]{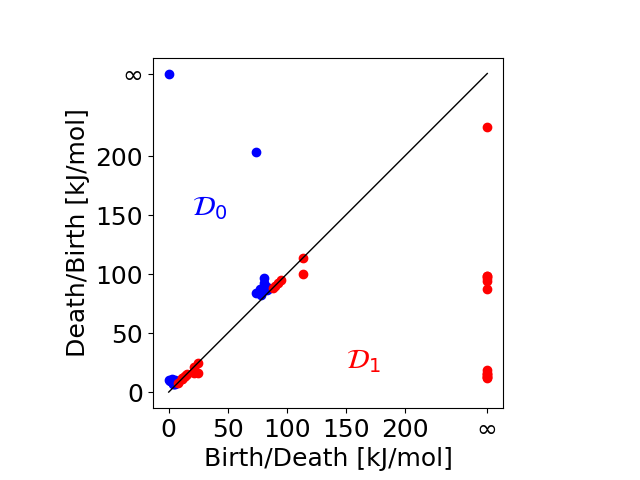}
  \subcaption{}
  \end{minipage}
  \caption{(a) Partial RRM of allyl vinyl ether C$_5$H$_8$O and (b) its persistence diagram}
  \label{fig:Claisen}
\end{figure*}

Finally, the proposed method is applied to partial RRM for the Claisen rearrangement ([3,3]-sigmatropic rearrangement) of allyl vinyl ether, C$_5$H$_8$O: 
\begin{figure}[H]
    \centering
    \includegraphics[keepaspectratio, scale=0.3]{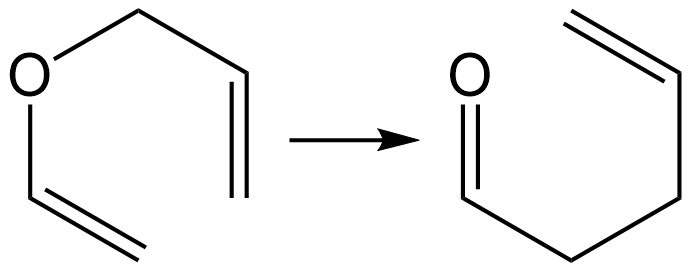}
\end{figure}
\noindent
which was reported in Ref.~\citenum{Nagahata2016_Claisen} and shown in Fig.~\ref{fig:Claisen}(a).
In this partial RRM, only [3,3]-sigmatropic rearrangements and conformational changes are included.
Figure \ref{fig:Claisen}(b) depicts the persistence diagram for this partial RRM.
In Fig.~\ref{fig:Claisen}(a), the RRM can be roughly divided into two regions (superbasins), namely, the left half, with high energy, corresponding to the reactant allyl vinyl ether structure and the right half, with low energy, corresponding to the product $\gamma, \delta$-unsaturated carbonyl structure.
Each region contains several EQ structures connected by TSs with low activation energies.
These two regions are connected by two elementary reaction paths with high activation energies.
In Fig.~\ref{fig:Claisen}(b), bundles of reactant and product regions appear around a diagonal line.
Namely, the product region in the RRM is represented by $0 \le b \le 25$ and $0 \le d \le 25$ in $\mathcal{D}_0$ and $\mathcal{D}_1$, as well as by $10 \le b \le 20$ and $d=\infty$ in $\mathcal{D}_1$.
However, the reactant region is represented by $70 \le b \le 115$ and $70 \le b \le 115$  in $\mathcal{D}_0$ and $\mathcal{D}_1$, as well as by $85 \le b \le 100$ and $d = \infty)$ in $\mathcal{D}_1$.
The plot at $(b,d)=(73.8,203.7)$ in $\mathcal{D}_0$ corresponds to the lowest-energy TS of the Claisen rearrangement.
While this TS connects the vertices with weights of 3.7 and 81.1, the reactant region constitutes the superbasin with the lowest weight of 73.8, leading to $b=73.8$.
The plot at $(b,d)=(224.4, \infty)$ in $\mathcal{D}_1$ indicates the existence of a second pathway connecting the reactant and product regions.
In summary, superbasins were displayed as plots around the diagonal line, as well as plots with similar $b$ but $d=\infty$ in $\mathcal{D}_1$, while the reactions connecting two superbasins were displayed as plots far away from the diagonal line.
As demonstrated here, the persistence diagram successfully captures the characteristics of the reaction pathways of the Claisen rearrangement. 

\section{Concluding Remarks}
\label{sec:Conclusion}
The GRRM program, which enables the automated search for reaction paths of chemical systems, extends the efficiency and applicability of quantum chemical calculations.
The program outputs RRMs, which consist of minima (EQs), first-order saddle points (TSs), and their connections (IRCs), to the PES of a chemical system. 
Most previous GRRM-based studies on the mechanisms of certain reactions have considered only a small portion of TSs with small activation energies and EQs connected by them. 
As the RRM of an $N$ atomic system contains all EQs and TSs, which are significantly important geometries among the $(3N-6)$-dimensional PESs from the perspective of chemistry, discarding information falling outside a small energy barrier is too wasteful.
Here, a computational method is proposed to extract the characteristic features of an RRM based on the PH applied to a weighted graph representing an RRM.
The proposed method provides a unique feature value called a persistence diagram or barcode for an RRM, which is independent of the depiction of the RRM.
In this method, the 0-th PH gives information about the connection between the EQ and TS, which corresponds to the analysis by the DG, while the 1-st PH gives information about the circumference of the reaction path.
Numerical demonstrations establish that these feature values reflect chemical information about the PES of the system, e.g., chemical hardness, element-pair affinity, and character of the reaction pathways.

If one is interested only in a particular reaction, a detailed analysis of the RRM of the system is sufficient.
However, the feature values extracted from the current scheme are expected to be useful for comparing or categorizing multiple RRMs.
For example, metal nanoclusters exhibit non-scalable catalytic activity depending on their size and composition.
Computational prediction of the catalytic activity of metal nanoclusters based on straightforward \textit{ab initio} calculations is often difficult because of their dependence on various external factors, such as temperature, support, and co-catalysts.
However, clusters with RRMs like those already known to be good catalysts can be expected to exhibit good catalytic activity.
Therefore, the utilization of the feature values of the persistence diagram as a descriptor of metal nanoclusters and further categorization of metal nanoclusters based on the similarity of RRMs is expected to be beneficial.
To achieve this, a method is required to measure the similarity between two persistence diagrams quantitatively.
Examples of such methods include the Wasserstein distance \cite{Tumer2014_PD-Wasserstein} and bottleneck distance \cite{Cohen-Steiner2007_StabilityPD}. However, the suitability of these metrics for this purpose remains unclear.
In addtion, as discussed in Sec.~\ref{sec:PH-sublevelset}, a general method for constructing a full network corresponding to $(3N-6)$-dimensional PES is expected to facilitate a deeper understanding of the system.
Another limitation of the current study is the absence of dissociation channels in RRMs.
Dissociation reactions represented as A $\to$ B + C are partially included in RRMs because molecules B and C often form so-called product complex.
However, molecular systems B and C also constitute independent RRMs at their dissociation limits.
Dissociation channels can be categorized into upward and downward channels, which can be searched optionally using the GRRM program.
The consideration of dissociation channels in RRMs and the hierarchical structure of RRMs in the current scheme is expected to broaden the applicability of the method and deepen the understanding extracted from it.

However, the adjusted weight rank clique filtration proposed here is applicable not only to RRMs but also to any type of weighted graph.
Because of the rapid increase in computer performance as well as the increased availability of network information, obtaining a vast network has become much easier than in the past.
The application of the proposed filtration to graphs other than those representing RRMs is expected to yield interesting results.

\acknowledgement

This work was supported in part by the Institute for Quantum Chemical Exploration (IQCE), JSPS KAKENHI for Transformative Research Areas ``Hyper-ordered Structures Science'' (Grant Numbers: JP21H05544 and JP23H04093 to M.K.), for Scientific Research (Grant Number: JP23H01915) and the Photo-excitonix Project of Hokkaido University.
We thank Prof.~Ohno (IQCE) for providing RRMs of organic molecules and Prof.~Maeda (Hokkaido Univ.) and Dr.~Nagahata (Hokkaido Univ.) for providing the RRM of the Claisen rearrangement.
Some of the reported calculations were performed using computer facilities at the Research Center for Computational Science, Okazaki (Projects: 21-IMS-C018 and 22-IMS-C019), and at the Research Institute for Information Technology, Kyushu University, Japan. 
The Institute for Chemical Reaction Design and Discovery (ICReDD) was established by the World Premier International Research Initiative (WPI) of MEXT, Japan.
We would like to thank Editage (www.editage.com) for English language editing.

\begin{suppinfo}

The Supporting Information is available free of charge via the Internet at http://pubs.acs.org.

\begin{itemize}
  \item One-to-one correspondences between the persistence barcode and RRM in Au$_5$
  \item Number of bars and statistics of bar lengths in the persistence barcodes
\end{itemize}

\end{suppinfo}

\bibliography{references}

\end{document}


\maketitle

\begin{figure}[H]
 \begin{minipage}[b]{\linewidth}
  \centering
  \includegraphics[keepaspectratio, scale=.8]
  {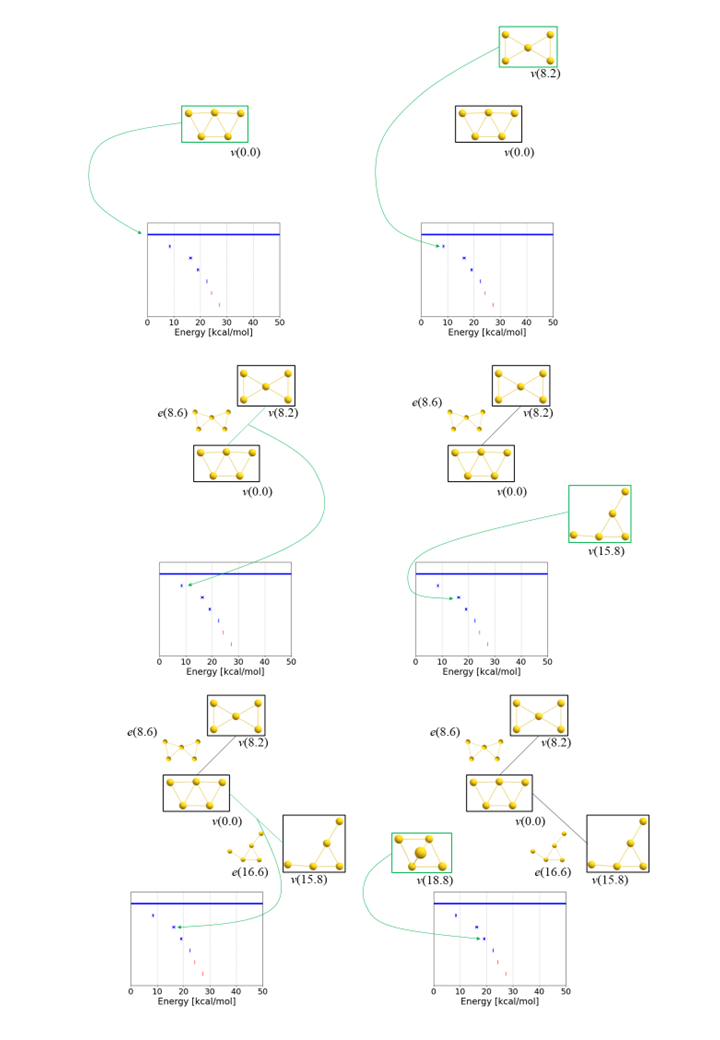}
  \label{Au5PBRRM1}
 \end{minipage}

 \label{fig:RRM_pentane}
\end{figure}

\begin{figure}

 \begin{minipage}[b]{\linewidth}
  \centering
  \includegraphics[keepaspectratio, scale=.8]
  {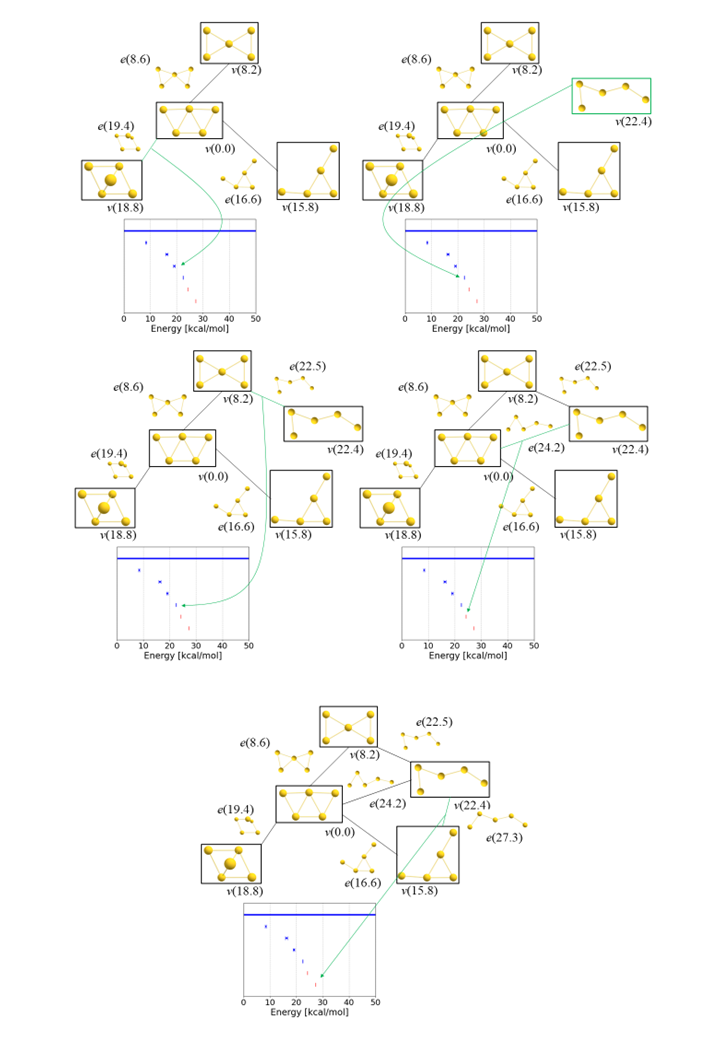}
  \label{Au5PBRRM2}
 \end{minipage}
 \caption{One-to-one correspondences between PB and RRM in Au$_5$}
 \label{fig:RRM_pentane}
\end{figure}

\begin{table}[h]
  \caption{Number of bars and average and standard deviation (SD) of bar lengths of PB of metals for $\mathcal{B}_0$ and $\mathcal{B}_1$}
  \label{table:data_simple_metal}
  \centering
  \begin{tabular}{lrrcrrcrr}
    \hline
    &\multicolumn{2}{c}{Number of bars} &
    &\multicolumn{2}{c}{Average lifetime} &
    &\multicolumn{2}{c}{SD of lifetime} \\
    \cline{2-3}
    \cline{5-6}
    \cline{8-9}
    metal & 0-th PH & 1-st PH && 0-th PH & 1-st PH && 0-th PH & 1-st PH \\
    \hline
    Au$_5$  & 5 & 2 && 0.479 & 0.000 && 0.235 & 0.000 \\
    Au$_7$  & 23& 26&& 0.897 & 0.562 && 1.181 & 2.198 \\
    Ag$_5$  & 3 & 0 && 0.649 &   -   && 0.000 & - \\
    Ag$_7$  & 9 & 9 && 1.188 & 0.638 && 1.295 & 1.804 \\
    Cu$_5$  & 2 & 0 && 2.032 &   -   && 0.000 & - \\
    Cu$_7$  & 6 & 2 && 0.837 & 0.000 && 0.901 & 0.000 \\
    Au$_4$Cu  & 8 & 3 && 0.732 & 0.0002 && 0.568 & 0.000\\
    Au$_3$Cu$_2$ & 20& 17&& 1.180 & 1.089  && 2.108 & 2.212\\
    Au$_2$Cu$_3$ & 13& 10&& 2.253 & 2.522  && 2.331 & 4.703\\
    AuCu$_4$  & 6 & 2 && 2.101 & 0.0002 && 1.696 & 0.000\\
    Au$_3$Ag$_2$  & 20& 18&& 1.339 & 0.356 && 2.587 & 1.179\\
    Au$_2$Ag$_3$  & 14& 9 && 2.591 & 1.791 && 2.635 & 2.992\\
    Ag$_3$Cu$_2$  & 8 & 8 && 3.253 & 1.924 && 2.322 & 4.231\\
    Ag$_2$Cu$_3$  & 9 & 9 && 3.509 & 1.638 && 3.010 & 2.931\\
    \hline
  \end{tabular}
\end{table}

\begin{table}[h]
  \caption{Number of bars and average and standard deviation (SD) of bar lengths of PB of organic molecules for $\mathcal{D}_0$ and $\mathcal{D}_1$}
  \label{table:data_organic}
  \centering
  \begin{tabular}{lrrrcrrcrr}
    \hline
    &\multicolumn{3}{c}{Number of bars} &
    &\multicolumn{2}{c}{Average lifetime} &
    &\multicolumn{2}{c}{SD of lifetime} \\
    \cline{2-4}
    \cline{6-7}
    \cline{9-10}
    compositions & 0-th PH & 1-st PH & 2-nd PH && 0-th PH & 1-st PH && 0-th PH & 1-st PH \\
    \hline
    H$_2$CO & 4 & 0 & 0 && 25.607 & - && 6.705 & - \\
    H$_4$C$_2$O$_2$ & 109 & 156 & 8 && 10.421 & 3.277 && 17.346 & 11.258 \\
    H$_4$C$_2$ & 2  & 0 & 0 && 3.303 & - && 0 & - \\
    H$_2$C$_2$N$_2$ & 63 & 93 & 4 && 25.209 & 1.126 && 24.114 & 3.541 \\
    HC$_2$NO$_2$ & 217 & 307 & 14 && 13.284 & 1.406 && 15.932 & 6.917\\
    C$_5$H$_8$O & 23 & 32 & 2 && 11.659 & 1.648 && 26.060 & 3.642\\
    \hline
  \end{tabular}
\end{table}